\documentclass[aps,preprintnumbers,amsmath,amssymb]{revtex4}
\usepackage{amsmath,amssymb,dsfont}
\usepackage{graphics,epsfig,mathtools}
\usepackage{amsfonts}
\usepackage{graphicx}
\usepackage{color,xfrac}
\usepackage{hyperref}
\usepackage{mathtools}
\usepackage{esint}
\usepackage{xcolor}
\usepackage[normalem]{ulem}
\usepackage{amsthm}
 \usepackage{booktabs}
 \usepackage{siunitx}
\theoremstyle{definition}

\theoremstyle{remark}

\newcommand{\Lt}{L_{\text{tot}}}

\def\build#1_#2^#3{\mathrel{\mathop{\kern 0pt#1}\limits_{#2}^{#3}}}
 \newcommand{\NS}{\mathrm{NS}}

\DeclareMathSymbol{\leqslant}{\mathalpha}{AMSa}{"36} 
\DeclareMathSymbol{\geqslant}{\mathalpha}{AMSa}{"3E} 
\DeclareMathSymbol{\eset}{\mathalpha}{AMSb}{"3F}     
\renewcommand{\leq}{\;\leqslant\;}                   
\renewcommand{\geq}{\;\geqslant\;}                   

\def\build#1_#2^#3{\mathrel{\mathop{\kern 0pt#1}\limits_{#2}^{#3}}}

\newcommand{\R}{\mathbb{R}}
\newcommand{\T}{\mathbb{T}}

\def \E{ \mathbb E  }

\begin{document}

\title{The spatio-temporal statistical structure of the turbulent dissipation field and its stochastic representation as a Gaussian Multiplicative Chaos}

\author{Wandrille Ruffenach$^{a}$ and  Laurent Chevillard$^{a,b,\dag}$}
\affiliation{
$^{\text{a}}$Univ Lyon, Ens de Lyon, Univ Claude Bernard, CNRS, Laboratoire de Physique, 46 all\'ee d’Italie F-69342 Lyon, France\\
$^{\text{b}}$CNRS, ICJ UMR5208, Ecole Centrale de Lyon, INSA Lyon, Universit\'e Claude Bernard Lyon 1, Université Jean Monnet, 69622 Villeurbanne, France\\
$^{\dag}$laurent.chevillard@cnrs.fr
}%

\begin{abstract}
The present article concerns the stochastic modeling of the turbulent dissipation field and in particular its temporal evolution. To do so, we will be calling for a random distribution, ubiquitous in several aspects of physics and probability theory, known as the Gaussian Multiplicative Chaos (GMC), that takes its roots in the phenomenology of fluid turbulence. Firstly introduced by Mandelbrot, shortly after Yaglom's discrete multiplicative cascade models, and rigorously studied by Kahane, the GMC appears as an appropriate statistically homogeneous model of the turbulent dissipation field. In this article, we will be recalling several ingredients of the associated turbulent phenomenology and its stochastic representation as a GMC, and propose a generalization to a spatio-temporal framework. All along the presentation of known properties in space, and in order to support new propositions concerning the temporal evolution, we will be calling for a comparison against Direct Numerical Simulations of the Navier-Stokes equations extracted from a publicly accessible database.  
\end{abstract}


\maketitle

\tableofcontents

\section{Introduction}
\label{s.intro}

Fluid turbulence is a natural phenomenon of prime importance in several fields of research in physics and engineering \cite{MonYag71,TenLum72,Fri95,Pop00}. Because of the interplay of the nonlinear and nonlocal nature of the underlying equations of motion, i.e. the Navier-Stokes equations, the understanding based on rigorous mathematics of the observed highly fluctuating nature of the velocity field, and related kinematic quantities of interest, is still in its infancy. Instead, a very precise and self-consistent phenomenology has been developed over the years \cite{Tay15,Ric22,Tay22,Kol41,Ons49,Kol62,Obu62,EyiSre06,Eyi24}.

We will be here focussing on the viscous dissipation field that plays the crucial role of transforming the mechanical energy, that is continuously injected through an external forcing term, into heat in a very efficient way. As we will be recalling in a precise way, its observed statistical signatures surprisingly turn out to barely
depend on viscosity. In the zero viscosity limit, the dissipation field behaves asymptotically in a distributional manner, and will be modeled in a very satisfactory way by a random field referred to as a Gaussian Multiplicative Chaos (GMC). The GMC, and its statistical structure, is the present center of interest of this article. Early introduced in the turbulence literature in the seventies, although already existing under various forms in quantum field theory \cite{Hoe71}, in the derivation of the partition function of the Coulomb gas in two dimensions \cite{DeuLav74,AlaJan81} (see the review \cite{Ser24})
and in random matrix theory \cite{Meh04,AndGui10,LamOst18,BerWeb18}, the GMC has been used extensively in solid state physics \cite{VerPon15,LakPon25}, mathematical finance and in the formulation of two-dimensional quantum gravity \cite{Nak04,DupShe11,RhoVar14,DavKup16,GuiRho19,BerPow25,ConTha26}.

As we will recall, as far as the phenomenology of fluid turbulence is concerned, the GMC is called for the stochastic representation of the viscous dissipation field, following the statistical constraints formalized by Kolmogorov and Obukhov \cite{Kol62,Obu62}. The first constructive approach consisting in building up the dissipation field as the multiplication of independent and positive random weights distributed over a dyadic tree is attributed to Yaglom \cite{Yag66}. Early understood to be intrinsically breaking statistical homogeneity (i.e. the invariance of the underlying probability law by translations), these discrete cascade models were then  recast in a statistically homogeneous formulation by Mandelbrot \cite{Man72,Man74} while proposing the so-called limit lognormal model, consisting in taking the exponential of a log-correlated Gaussian random field. Kahane \cite{Kah85} then gave a proper mathematical meaning to this random field following a regularizing procedure, computed its statistical properties and renamed this probabilistic object in the limit as a GMC.

The article is organized as follows. In Section \ref{Sec:PhenoTurbulence}, we recall several important ingredients of the phenomenology of fluid turbulence, with a particular focus on the dissipation field. In particular, we recall the statistical constraints on the coarse-grained dissipation field over a ball in §\ref{Sec:KO62MomentsDiss} formulated by 
Kolmogorov and Obukhov to explain the observed intermittent behavior of the velocity field, and that led Yaglom to infer the logarithmic correlation structure of the logarithm of the dissipation, as precisely stated in §\ref{Sec:StatStructDiss}. We present then the analysis in §\ref{Sec:NumObservationsHopkins} of the Johns Hopkins turbulence database where are gathered various direct numerical simulations of the forced Navier-Stokes equations. This analysis allows us to observe the aforementioned peculiar correlation structure of the logarithm of dissipation, and moreover explore the correlation structure in time. To reproduce these behaviors, we begin with recalling key theoretical ingredients of the GMC in §\ref{Sec:SpatialGMC}, with a particular emphasis on its relevance to give a stochastic representation of the phenomenology previously depicted. Then, in §\ref{Sec:SpatioTempGMC}, we propose an extension of these considerations to a spatio-temporal framework. The presentation of the numerical method allowing to simulate instances of the spatio-temporal GMC and a final comparison against DNS of the Navier-Stokes equations are then proposed in §\ref{Sec:SimSTGMCandDNS}. Conclusions and perspectives are then gathered in §\ref{Sec:Conclusion}.

\section{A glimpse at the phenomenology of fluid turbulence}\label{Sec:PhenoTurbulence}

Let us begin by explaining why and how the GMC emerges from the physics of fluid turbulence. To do so, we first need to present the axiomatic approach, mainly attributed to Kolmogorov \cite{Kol41,Kol62}, although his work follows the ones of many people. The reader should find many references and historical origins in the book by Frisch \cite{Fri95}.

\subsection{The Navier-Stokes equations and the external stirring force}

Fluid turbulence has an ancient tradition in fluid mechanics and has been widely studied from an experimental point of view \cite{ComCor66,MonYag71,TenLum72} (see also for instance \cite{AnsGag84,CasGag90,SreAnt97,ArnBau96,ChaCha00,Tsi01,Wal09,BodBew14} to cite a few). Today, when focussing on the simplest situation of homogeneous and isotropic flows, the statistical structure of fluid turbulence is observed in numerical simulations of the Navier-Stokes equations  \cite{OrsPat72,Ker85,EswPop88,YeuPop89,VinMen91,LiPer08,IshGot09,YeuRav25}, usually assuming periodic boundary conditions and relying massively on parallel algorithms of the Fast Fourier Transform (FFT). Considering an incompressible (i.e. divergence-free) velocity field $u^\nu (t,x)\in \R^3$, which depends implicitly on the kinematic viscosity $\nu>0$, for $t\ge 0$ and $x\in\R^3$, these nonlinear and nonlocal partial differential equations read
\begin{align}\label{eq:NSforced}
\frac{\partial u^\nu}{\partial t} + (u^\nu\cdot \nabla)u^\nu = -\nabla p^\nu +\nu \Delta u^\nu +f_L,
\end{align}
where $p^\nu(t,x)$ is the pressure field and $f_L(t,x)\in \R^3$ an external forcing field that we will define later. Given an initial condition $u^\nu(0,x)$ that we will take to be vanishing without loss of generality, the evolution given in \eqref{eq:NSforced} is closed when assuming incompressibility, which allows to express pressure $p^\nu(t,x)$ as a function of the velocity field through Poisson's equation \cite{MajBer02,FoiMan01}.

Concerning the physics of fluid turbulence, the external additional forcing field $f_L(t,x)\in \R^3$ entering in  \eqref{eq:NSforced} is of crucial importance. It should be viewed as a schematic model of the action of a stirring procedure performed at large scales by propellers, as it happens in the emblematic Von K\'arm\'an Swirling Flows \cite{ZanDij87,DouCou91,RavChi08,SaiDub14}, or stemming from the instability of fluid layers when going through a grid in a wind tunnel (see for instance the historical articles \cite{BatTow48a,BatTow48b,ComCor66}). Its role is clear and is devoted to keeping injecting energy into the system that will eventually be transformed into heat by the viscous term. It is required to be smooth in space for physical reasons. Smoothness-in-space of the forcing term $f_L$ is then guaranteed by populating Fourier modes over a finite range of wave vectors, and more precisely in the vicinity of the wave-vectors of amplitude of order $1/L$. The characteristic length scale $L$ will eventually play a crucial role in the phenomenology of fluid turbulence, and will govern the correlation length scale of the velocity field. About the forcing $f_L$, besides being smooth-in-space, its precise nature has little importance as it has been extensively studied in the literature. In particular, it has been observed that a deterministic and constant-in-time version of the forcing as proposed in \cite{VinMen91}, or a random version, no matter how the temporal structure is chosen (i.e. independent instances or finitely correlated in time), as proposed in \cite{EswPop88}, give a very similar statistical picture at \textit{small scales}, i.e. for the velocity increments and velocity gradients. This universal behavior of the velocity field with respect to the very nature of the forcing illustrates the robustness of the phenomenology that we will be recalling. In the following, we will be considering for example a random forcing and forthcoming expectations will be taken over its instances.

\subsection{Establishment of a statistically stationary regime and the anomalous behavior of velocity variance and average dissipation}

As time goes on, starting for instance from a vanishing initial condition $u^\nu(0,x)=0$, it is observed that velocity reaches a statistically homogeneous, isotropic and stationary regime, with corresponding invariance of the underlying probability law by spatial translations, rotations and temporal translations. Furthermore, in this regime obtained after a transient of finite duration determined by the precise shape and amplitude of the forcing term $f_L$, velocity is of zero average, as it is expected from statistical isotropy, and its variance turns out to depend only weakly on viscosity. We ultimately obtain
\begin{align}\label{eq:FiniteVarNuTo0}
\lim_{\nu\to 0}\lim_{t\to\infty}\E\left|u^\nu(t,x)\right|^2\equiv\sigma^2<+\infty,
\end{align}
where the expectation is taken over the instances of the forcing. The limiting behavior of the velocity variance with viscosity \eqref{eq:FiniteVarNuTo0} is indeed surprising, and says that fluids are able to dissipate energy in a very efficient way, such that in particular the average kinetic energy remains finite despite keeping injecting energy into the dynamics through the forcing term $f_L$ \eqref{eq:NSforced}. For instance, the respective stochastic heat equation obtained when discarding the nonlinear term and pressure gradient from  \eqref{eq:NSforced} would instead generate a statistically stationary regime in which the average dissipation
rate remains finite, but with a velocity variance that would instead be proportional to $\nu^{-1}$ \cite{BecBre24}. In order to warrant such an efficient way to dissipate energy, the fluid will develop \textit{small scales} following a cascading process of energy through scales. As a consequence, velocity Fourier modes  at wave numbers much larger than those excited by the forcing term will be populated according to the ``$k^{-5/3}$"-Kolmogorov's power spectral density. Correspondingly, velocity gradients will reach very \textit{high} values, such that the viscous term entering in the dynamics \eqref{eq:NSforced}, made up of the velocity second derivatives, gets exceptionally large.

The development of high amplitude of the gradients is at the core of the phenomenology of fluid turbulence. Very early \cite{Tay15,Ric22,Tay22}, the focus has been put on the fluctuations of the so-called dissipation field $\varepsilon^\nu(t,x)$ which is defined by
\begin{align}\label{eq:DefDissField}
\varepsilon^\nu(t,x) = \frac{\nu}{2} \sum_{i,j=1}^{3}\left( \frac{\partial u^\nu_i}{\partial x_j}+\frac{\partial u^\nu_j}{\partial x_i}\right)^2,
\end{align}
where the velocity field $u^\nu$ is the solution of the forced Navier-Stokes equations \eqref{eq:NSforced}. The average of this field plays a key role in the kinetic energy budget, which corresponds to the time evolution of the average of the kinetic energy $|u^\nu(t,x)|^2$. As surprising as is the observation of the finiteness of the velocity fluctuation variance as $\nu\to 0$ \eqref{eq:FiniteVarNuTo0}, and intimately related,  the average dissipation $\overline{\varepsilon}$, i.e. the expectation of $\varepsilon^\nu(t,x)$ \eqref{eq:DefDissField}, remains also finite and non vanishing in the same limit \cite{Fri95,SreAnt97,KanIsh03,ValTha24}:
\begin{align}\label{eq:FinitAverageDissField}
0<\overline{\varepsilon}\equiv \lim_{\nu\to 0}\lim_{t\to\infty}\E\left[\varepsilon^\nu(t,x)\right] <\infty.
\end{align}
The finiteness of $\overline{\varepsilon}$ \eqref{eq:FinitAverageDissField} is known as the \textit{anomalous dissipation} in turbulence literature. Roughly speaking, the asymptotic behavior proposed in \eqref{eq:FinitAverageDissField} says that that velocity gradients variance diverges as $\nu^{-1}$ as viscosity goes to zero. In this limit, we are left with the single possibility that $\overline{\varepsilon}$ \eqref{eq:FinitAverageDissField} is related to velocity variance $\sigma^2$ \eqref{eq:FiniteVarNuTo0} according to the dimensional relation
\begin{align}\label{eq:FinitAverageDissFieldToSigma}
\overline{\varepsilon}\propto \frac{\sigma^3}{L},
\end{align}
where $L$ is the characteristic correlation length of the forcing term $f_L$ entering in \eqref{eq:NSforced}, the remaining multiplicative constant that would enter in \eqref{eq:FinitAverageDissFieldToSigma} being universal, in particular independent of the precise shape of the flow at large scales and of the nature of the forcing $f_L$.  These surprising anomalous behaviors of kinetic energy \eqref{eq:FiniteVarNuTo0}  and average dissipation \eqref{eq:FinitAverageDissField} are the building blocks of the statistical physics of fluid turbulence, and have been first formulated more than a century ago \cite{Tay15,Tay22}, as reviewed in \cite{EyiGol25}.

\subsection{The moments of the locally averaged dissipation field}\label{Sec:KO62MomentsDiss}

As we have seen, the dissipation field plays a key role in the statistical structure of the turbulent velocity field. As a random field, it is expected to be a positive quantity of a given average, determined by the structure of the flow \eqref{eq:FinitAverageDissField}, in particular by the velocity variance $\sigma^2$ and the so-called integral length scale $L$, even in the asymptotic limit of vanishing viscosity. We could now wonder how  higher-order moments of the dissipation field behave. Making a long story short, in order to interpret extreme events of the field of velocity gradients, as they were early observed in wind tunnels  \cite{BatTow48a,BatTow48b,ComCor66}, and nowadays referred to as the intermittency phenomenon, Kolmogorov and Obukhov \cite{Kol62,Obu62} jointly proposed to assume $\varepsilon^\nu(t,x)$ being distributional in nature in the limit $\nu\to 0$. As such, they proposed, instead of studying the dissipation field itself,  to consider a locally averaged version of the dissipation field over a ball of radius $\ell$, say $\varepsilon_\ell^\nu(t,x)$, obtained as the following convolution
\begin{align}\label{eq:DefDissFieldOverBall}
\varepsilon^\nu_\ell (t,x) \equiv \int_{\R^3}\varphi_\ell(x-y)\varepsilon^\nu(t,y) dy,
\end{align}
where we have introduced a dilated version $\varphi_\ell(x)=\varphi(x/\ell)/\ell^3$ of a unit-volume ball of unit-radius, i.e.
\begin{align}\label{eq:TestFunctionBall}
\varphi(x)=\frac{3}{4\pi}\mathds{1}_{|x|\le 1}.
\end{align}

Doing so, they were able to interpret the intermittent, i.e. anomalous, corrections observed on the velocity field using a dimensional argument based on the coarse-grained dissipation field \eqref{eq:DefDissFieldOverBall} instead of the dissipation field itself, known as the Refined Similarity hypothesis (RSH). This procedure and its historical origin are especially clearly exposed in \cite{Man72} and reviewed in \cite{Fri95}, and allows to generalize the asymptotic behavior of the average dissipation \eqref{eq:FinitAverageDissField} to higher-order moments as
\begin{align}\label{eq:FinitHOMomentsDissField}
 \lim_{\nu\to 0}\lim_{t\to\infty}\E\left[ \left(\varepsilon^\nu_\ell\right)^q\right] \build{\sim}_{\ell\to 0}^{} c_q\overline{\varepsilon}^q \left( \frac{\ell}{L}\right)^{\tau_q},
\end{align}
where $\tau_q$ is the spectrum of exponents and is asked to be a non-linear function of the order $q$ such that $\tau_1=0$, and $c_q$ are universal pre-factors such that $c_1=1$. In current language, the coarse-grained version of the dissipation field \eqref{eq:DefDissFieldOverBall} can be seen as a mollifying procedure obtained while integrating the local dissipation against a test function $\varphi_\ell$ seen as an approximation of the Dirac-$\delta$ function (here a ball of radius $\ell$ properly normalized by its volume), such that $\varepsilon^\nu_\ell (t,x)\to \varepsilon^\nu (t,x)$ as $\ell\to 0$, at each position $x$ and at any time $t$. The spectrum of exponents $\tau_q$ is called the multifractal spectrum \cite{Fri95}. In order to make this approach consistent with experimental observations, for reasonable values of the  order $q$, the authors of \cite{Kol62,Obu62} furthermore propose the simple quadratic model
\begin{align}\label{eq:LNTAUQ}
\tau_q = -\mu\frac{q(q-1)}{2},
\end{align}
where $\mu>0$ is called the intermittency coefficient in the turbulence literature. Such a quadratic approximation of the multifractal spectrum $\tau_q$ \eqref{eq:LNTAUQ} is known as the lognormal model, for reasons that will become clear later, and has been shown to be a fairly good representation of experimental \cite{SreAnt97} and numerical (see \cite{BuaSre22} for a recent analysis) data for moderate orders $1\le q\le 4$, with the observed universal coefficient $\mu=0.2\pm 0.02$ \cite{AntPha81,AntSat82,TanAnt20}.

\subsection{The spatial statistical structure of the dissipation field}\label{Sec:StatStructDiss}

After the work of Kolmogorov and Obukhov \cite{Kol62,Obu62}, it became clear that the fluctuating nature of the dissipation field in the limit of vanishing viscosities should be considered in a distributional fashion. In particular, notice that whereas the average $\overline{\varepsilon}$ is expected to be finite \eqref{eq:FinitAverageDissField}, its second moment should diverge. This can be clearly seen when using the proposition made in \eqref{eq:FinitHOMomentsDissField}: we indeed expect a divergence of the second moment $\E [ \left(\varepsilon^\nu_\ell\right)^2]$ as $\ell^{-\mu}$ in the inviscid limit $\nu\to 0$, when the scale of the mollifier $\ell$ \eqref{eq:DefDissFieldOverBall} goes to zero. Especially emphasized in the work of Obukhov \cite{Obu62}, it will be furthermore assumed that the density of $\varepsilon^\nu$ is lognormal, meaning that its logarithm is Gaussian. 

We will note
\begin{align}\label{eq:DefLNLogDiss}
\frac{\ln \varepsilon^\nu(t,x)- \E\left[\ln \varepsilon^\nu\right]}{\sqrt{\mathbb{V} \left[\ln \varepsilon^\nu\right] }} \build{=}_{}^{\text{law}} \mathcal N \left(0,1\right),
\end{align}
where the former equality is true \textit{in law}, i.e. relating an equality of the respective one-point and one-time probability densities of the left and right hand sides, and $\mathcal N(0,1)$ the standard normal (Gaussian) random variable of zero average and unit variance.

Experimental investigations discussed in \cite{GurZub63}, reviewed and updated in \cite{TanAnt20}, demonstrated that the dissipation is furthermore correlated over the largest scale of the flow, i.e. over a scale of the order of the characteristic length scale of the forcing $L$. This is indeed very surprising for several reasons since it is expected naively that velocity gradients should be correlated over a small dissipative (i.e. governed by viscosity) length scale. This turns out to be true for gradients, but not their amplitude, as it is involved when squaring them \eqref{eq:DefDissField}. This long-range correlated nature of the gradients amplitude and the anomalous scaling of the moments of the coarse-grained dissipation field \eqref{eq:FinitHOMomentsDissField}, with a quadratic multifractal spectrum \eqref{eq:LNTAUQ}, were unified by the theoretical proposition of Yaglom \cite{Yag66}. In his stochastic framework, the dissipation field is the result of a discrete multiplicative process, in which multipliers are distributed over a dyadic structure. The resulting random field, more generally referred in the literature as Mandelbrot's discrete multiplicative cascades \cite{Man72,Man74,KahPey76}, is at the root of early investigations on the stochastic modeling of fluid turbulence, including \cite{MenSre87,MenSre91,BenBif93,ArnBac98a} to cite a few.

Defining the dissipation field as the product of independent positive multipliers distributed over a dyadic decomposition of space, as it is especially clearly defined and reviewed in \cite{Mol96,ArnBac98a}, has a clear consequence on its spatial structure. In the following, we use $\ell$ to denote either a vector in $\R^d$, its norm, or a scalar argument, with the meaning clear from context. Assuming a statistically homogeneous and stationary framework, let us define then the correlation function of the logarithm of the dissipation field as
\begin{align}\label{eq:DefCorrLogDiss}
\mathcal C_{\ln \varepsilon^\nu } (\ell) =\E\left[ \ln \varepsilon^\nu (t,x) \ln\varepsilon^\nu (t,x+\ell)\right]_c,
 \end{align}
where the subscript $c$ entering in the expectation \eqref{eq:DefCorrLogDiss} says that respective averages have been subtracted in the definition the correlation function, i.e. $\E[xy]_c=\E[xy]-\E[x]\E[y]$. Early understood in  \cite{Yag66}, it was predicted that the dissipation field $\varepsilon^\nu (t,x)$ \eqref{eq:DefDissField}, in the limit of vanishing viscosities, would be correlated according to
\begin{align}\label{eq:CorrLogYaglom}
\mathcal C_{\ln \varepsilon } (\ell) = \lim_{\nu\to 0}\mathcal C_{\ln \varepsilon^\nu }(\ell)  = \mu \ln_+\left( \frac{L}{|\ell|}\right) + \mu f_{\NS}(\ell),
 \end{align}
where $f_\NS$ is a bounded and continuous function of its argument, $\ln_+(|x|)=\max\left(\ln(|x|),0\right)$ and $\mu>0$ a free parameter of this construction related to the random nature of the multipliers. Once again, let us mention that the construction of Yaglom \cite{Yag66} is not statistically homogeneous, and as a consequence, the correlation proposed in \eqref{eq:CorrLogYaglom} should also depend on the very position $x$. As it is proposed in \cite{ArnBac98a}, a way to get rid of the dependence on the positions $x$ is to consider an empirical average version while performing an appropriate sum over them, that allows to compute explicitly the corresponding bounded function $f_\NS$ entering in 
\eqref{eq:CorrLogYaglom}. 

When dealing with random continuous fields, the one-point one-time lognormality depicted in \eqref{eq:DefLNLogDiss} should be considered in a broader sense. We will assume then that the random vector made of values taken by $\ln \varepsilon^\nu (t,x_i)$ over any finite set of positions $x_i$ is jointly normal, with covariance matrix given $\mathcal C_{\ln \varepsilon^\nu }(x_i-x_j)$. This says that any linear combination of the components of such a vector is a Gaussian random variable, and gives a complete characterization of the dissipation field and its one-time and $q$-points correlator which reads, for any integer $q\ge 2$,
\begin{align}
\E\left(\prod_{i=1}^q\varepsilon^\nu (t,y_i)\right) = \E\left(e^{\sum_{i=1}^q\ln\varepsilon^\nu (t,y_i)}\right) &= e^{q \E\left[ \ln \varepsilon^\nu \right]} \E\left(e^{\sum_{i=1}^q\left(\ln\varepsilon^\nu (t,y_i)- \E\left[ \ln \varepsilon^\nu \right]\right)}\right)\notag\\
&=e^{q \E\left[ \ln \varepsilon^\nu \right]} e^{\frac{1}{2}\sum_{i,j=1}^q\mathcal C_{\ln \varepsilon^\nu}(y_i-y_j)}\notag\\
&=\left(\E \left[ \varepsilon^\nu\right]\right)^qe^{\sum_{i<j=1}^q\mathcal C_{\ln \varepsilon^\nu}(y_i-y_j)}\label{eq:CorrelatorsDiss},
  \end{align}
where we have used the fact that $\E \exp (g) = \exp(\mathbb{V}[g]/2)$ for any zero-average Gaussian random variable $g$, and in particular
\begin{align}\label{eq:ConsLNLogDissAverageVar}
\E \left[ \varepsilon^\nu\right]= e^{\E\left[ \ln \varepsilon^\nu \right]+\frac{1}{2}\mathbb{V} \left[\ln \varepsilon^\nu\right]}.
\end{align}

It is then straightforward to make a connection between the correlation structure of the dissipation field \eqref{eq:CorrLogYaglom} in the asymptotic limit $\nu\to 0$ and the behavior of the moments  \eqref{eq:FinitHOMomentsDissField} of its coarse-grained version  \eqref{eq:DefDissFieldOverBall}. To see this, following a similar chain of developments as they are proposed in \eqref{eq:CorrelatorsDiss}, using the expression of the average dissipation \eqref{eq:ConsLNLogDissAverageVar}, we obtain for any integer $q\ge 2$
\begin{align}\label{eq:DerivCGDissSmallScales}
\E\left[ \left(\varepsilon^\nu_\ell\right)^q\right] &=  \E \int_{(\R^3)^q}\prod_{i=1}^q\varphi_\ell(x-y_i)\varepsilon^\nu (t,y_i)dy_i =\left(\E \left[ \varepsilon^\nu\right]\right)^q\int_{(\R^3)^q}\left(\prod_{i=1}^q\varphi_\ell(x-y_i)\right) e^{\sum_{i<j=1}^q\mathcal C_{\ln \varepsilon^\nu}(y_i-y_j)}\prod_{i=1}^q dy_i.
\end{align}
Let us now take the limit $\nu\to 0$ in the former expression \eqref{eq:DerivCGDissSmallScales}. Using the notation of \eqref{eq:FinitAverageDissField} and the asymptotic form of the covariance \eqref{eq:CorrLogYaglom}, commuting in a formal way the integral and the limit, we obtain
\begin{align}
 \lim_{\nu\to 0}\E\left[ \left(\varepsilon^\nu_\ell\right)^q\right]  &=\overline{\varepsilon}^q\int_{(\R^3)^q}\left(\prod_{i=1}^q\varphi_\ell(y_i)\right) \prod_{i<j=1}^q \left| \frac{L}{y_i-y_j}\right|_+^\mu e^{\mu\sum_{i<j=1}^q  f_\NS(|y_i-y_j|)}\prod_{i=1}^q dy_i\notag\\
 &=\overline{\varepsilon}^q\int_{(\R^3)^q}\left(\prod_{i=1}^q\varphi(y_i)\right) \prod_{i<j=1}^q \left| \frac{L}{\ell(y_i-y_j)}\right|_+^\mu e^{\mu\sum_{i<j=1}^q  f_\NS(\ell |y_i-y_j|)}\prod_{i=1}^q dy_i\notag\\
 &\build{\sim}_{\ell\to 0}^{}c_q\overline{\varepsilon}^q \left( \frac{\ell}{L}\right)^{\tau_q},\label{eq:PowerLawMoments}
\end{align}
where $|\cdot|_+=\exp(\ln_+|\cdot|)$, $\tau_q$ given in \eqref{eq:LNTAUQ} and
\begin{align}\label{eq:ExpCqLN}
c_q = e^{\mu\frac{q(q-1)}{2}f_\NS(0)}\left(\frac{3}{4\pi}\right)^q\int_{|y_i|\le 1}\prod_{i<j=1}^q \frac{1}{|y_i-y_j|^\mu}\prod_{i=1}^q dy_i.
\end{align}

Determining the range of possible values for the intermittency coefficient $\mu$ and the order $q$ ensuring that the multiplicative factor $c_q$ \eqref{eq:ExpCqLN} remains finite is a tricky question when attacked from scratch. The multiple integral entering in the multiplicative constant is a generalization to high dimensions of Selberg's integral \cite{ForWar08} and appears in several fields, such as log-gases and random matrices \cite{For10,Lew22,Ser24}, and requires subtle analysis to get the sharp condition
\begin{align}\label{eq:SharpCondCq}
c_q <\infty \Leftrightarrow \mu<\frac{6}{q}.
\end{align}
A early demonstration of the condition \eqref{eq:SharpCondCq} in any space dimension $d$ (the condition becomes $\mu<2d/q$) can be found in \cite{Kah85} and \cite{Ros87}. Permutation of limits and integrals can then be justified using dominated convergence arguments, as they are presented in \cite{Kah85,RhoVar14}.

Let us recall that the phenomenological picture of Kolmogorov and Obukhov \cite{Kol62,Obu62} concerns the limiting power-law behavior of the moments of order $q$ of the coarse-grained dissipation field \eqref{eq:FinitHOMomentsDissField} with a quadratic spectrum of exponents \eqref{eq:LNTAUQ} of parameter $\mu$. Former considerations show that, up to the moment of order of $q$, these prescriptions on the statistical structure of the dissipation field can be fulfilled by a lognormal random field \eqref{eq:DefLNLogDiss}, fully characterized by the logarithmic correlation structure given in \eqref{eq:CorrLogYaglom}, as long as $\mu$ is smaller than a $q$-dependent bound \eqref{eq:SharpCondCq}. This being said, given the value of the intermittency parameter $\mu\approx 0.2$, these aforementioned assumptions give a consistent picture up to the order $6/\mu\approx 30$, which is much larger than what can be estimated on empirical data with enough statistical confidence.

The very purpose of the Gaussian Multiplicative Chaos (GMC) theory is to give a meaning to such a lognormal field, in a distributional sense, and propose a construction based on a limiting procedure.

\subsection{Comparison against Direct Numerical Simulations of the Navier-Stokes equations}\label{Sec:NumObservationsHopkins}

Direct numerical simulations (DNSs) of the Navier-Stokes equations \eqref{eq:NSforced} constitute a remarkably robust tool to investigate the statistical properties of turbulent flows. We will now confront the different expected statistical behaviors exposed above with the publicly available database of DNSs of the Johns Hopkins university \cite{LiPer08}.

When comparing against empirical (experimental or numerical) data, we need to introduce finite-viscosity, or equivalently finite Reynolds number effects. The Reynolds number $\mathcal R_e$ is given by the following non dimensional number 
\begin{align}\label{eq:DefRe}
\mathcal R_e\equiv \frac{\sigma L}{\nu},
\end{align}
where $\sigma^2$ is the velocity variance that becomes independent of viscosity $\nu$ in the inviscid limit \eqref{eq:FiniteVarNuTo0}, and $L$ the characteristic length scale of the forcing $f_L$ \eqref{eq:NSforced}. At a finite Reynolds number, or equivalently for a finite viscosity, the asymptotic behaviors displayed in \eqref{eq:FinitHOMomentsDissField} and \eqref{eq:CorrLogYaglom} are expected only over the finite range of scales $\eta_K\ll \ell\ll L$, called the \textit{inertial range} in the turbulence literature \cite{TenLum72,Fri95,Pop00}, where enters Kolmogorov's dissipative length scale $\eta_K$, estimated on dimensional ground as
\begin{align}\label{eq:EtaK}
\eta_K \propto \left( \frac{\nu^3}{\overline{\varepsilon}}\right)^{1/4}\propto L\mathcal R_e^{-3/4},
\end{align}
 at which viscosity starts acting while regularizing the rough behaviors formerly depicted and making velocity gradients finite. 
 
The fully de-aliased simulations of \eqref{eq:NSforced} made available in \cite{LiPer08} are run in a three dimensional cube of side length $2\pi$ and provide turbulent fields in a stationary, homogeneous and isotropic setting. The database contains simulations with $N=1024^3$, $4096^3$, $8192^3$, and $32,768^3$ collocation points, corresponding to several increasing Reynolds numbers $\mathcal R_e\simeq 6.8\times 10^3, \, 1.4\times 10^4, \, 7\times 10^4$, and $1.8\times 10^5$. The analysis is carried out in two stages. 

We first examine the one-point statistics and the spatial correlations of $\ln \varepsilon^\nu$. Exploiting statistical homogeneity and isotropy, we estimate the spatial statistics on $N_s=20$ equally spaced two-dimensional slices in each coordinate direction, yielding a total of $3N_s$ slices for each dataset. For each simulation, we query both the velocity and velocity gradient fields from the database in order to estimate velocity related quantity such as Reynolds number and gradient related quantities such as the dissipation field. For the $N=1024^3$ simulation a temporal evolution of DNS fields on $N_t=5028$ time steps distant from $\delta t = 0.02$ time units is furthermore available. In the case of the $1024^3$ DNS, statistics are further averaged over time. This study of the one point statistics and correlations of the logarithm of the dissipation field allows us to retrieve existing results concerning the spatial structure of the energy dissipation rate.

Then, we also investigate the temporal correlations of $\ln \varepsilon^\nu$ and compare them with the spatial ones. This temporal analysis is restricted to the $N=1024^3$ DNS, since the other datasets do not provide time-resolved fields. Let us mention that an early investigation of the temporal structure of the logarithm of the dissipation can be found in \cite{PopChe90}, although at that time the Reynolds number associated to the simulation was much smaller.

\begin{figure}[htb]
	\centering
	\includegraphics[width=0.85\linewidth]{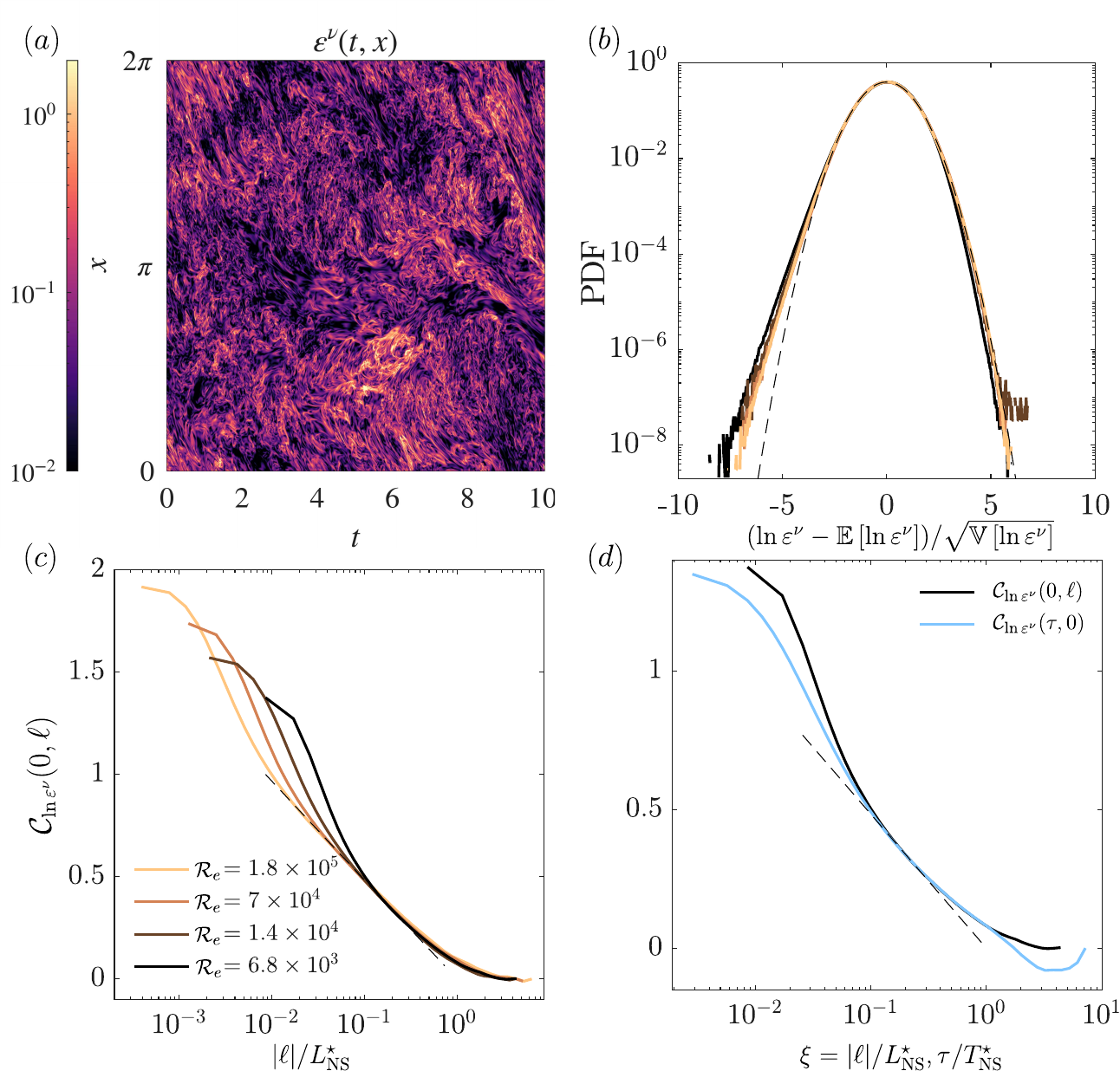}
	\caption{  (a) Spatio-temporal representation of the dissipation field $\varepsilon^\nu$ along the line $(x,0,0)$ extracted from the $N=1024^3$ DNS. The logarithmic colorbar unveils the large scale correlations in both space and time of turbulent dissipation. (b) Estimation of the unit variance one point statistics of the logarithm of dissipation at varying Reynolds number (solid lines). The color code is the same as in (c) and corresponds to each value of $\mathcal{R}_e$. The black dashed line is the unit variance and zero mean Gaussian probability density for comparison.(c) Estimation from the DNSs data of the spatial correlations of $\ln \varepsilon^\nu$ at different Reynolds numbers (solid lines) as a function of $|\ell|/L^\star_\NS$. The black dashed line represents the scaling $\mu \ln (L^\star_\NS/|\ell|)$ with $\mu=0.21$. (d) Comparison between spatial (solid black line) and temporal (solid blue line) correlations of $\ln\varepsilon^\nu$ estimated from the DNS data at $N=1024^3$ as a function of $|\ell|/L_\NS^\star$ in the fully spatial case and $\tau/T_\NS^\star$ in the temporal case. The black dashed line represents $\mu \ln (1/\xi)$ with $\mu=0.21$ and $\xi = \ell /L^\star_\NS, \tau/T^\star_\NS$.  }\label{fig:DNS}
\end{figure}

The corresponding results are displayed in Figure~\ref{fig:DNS}. Panel (a), shown for illustration, represents the evolution over all available timesteps of the dissipation field $\varepsilon^\nu$ of the $1024^3$ DNS along the line $(x,0,0)$ in a logarithmic color scale; it already suggests the presence of large scale structures both in space and in time. One can additionally note the filamentary aspect  of these large scales, reminiscent of the so-called sweeping effect as it is discussed in \cite{ChaLem26}.

Panel (b) shows the one point probability density of the normalized variable \eqref{eq:DefLNLogDiss}
estimated from histograms. The solid curves correspond to the different Reynolds numbers, ranging from black ($N=1024^3$) to yellow ($N=32,768^3$), while the black dashed curve indicates the standard Gaussian density. Two observations emerge. Firstly, once centered and normalized, the distribution of $\ln \varepsilon^\nu$ depends only weakly on the Reynolds number while both its average and variance are expected to diverge with the Reynolds number. Secondly, the normalized distributions remains close to a Gaussian shape, although departing from it for the lowest values of the dissipation. These observations are consistent with early analysis of DNSs, as they can be found in \cite{YeuPop89}. As noticed in \cite{YeuPop89}, and further exploited in \cite{PopChe90}, whereas the dissipation field $\varepsilon^\nu$ \eqref{eq:DefDissField} departs from log-normality for the lowest fluctuations, an associated field called pseudo-dissipation, defined as the Frobenius norm of the velocity gradient tensor \cite{Pop00} and that shares several statistical properties with the dissipation field, is observed to be very close to a log-normal random variable (see \cite{LetGou21} for a recent discussion on this subject). As far as we are concerned, we will nonetheless consider that log-normality is a good approximation of the density of the logarithm of dissipation, and as a consequence, the spatial and temporal statistical structure of the dissipation field will be correctly captured by the covariance  of $\ln \varepsilon^\nu$.

We therefore consider the corresponding space-time correlation function $\mathcal{C}_{\ln \varepsilon^\nu}(\tau,\ell)$, as a generalization of the formerly defined purely spatial correlation of  \eqref{eq:DefCorrLogDiss}, that reads in the statistically homogeneous and stationary regime
\begin{equation}\label{eq:DefCorrSTLogDiss}
	\mathcal{C}_{\ln \varepsilon^\nu}(\tau,\ell)= \mathbb{E}\left[\ln \varepsilon^\nu(t,x)  \ln \varepsilon^\nu(t+\tau,x+\ell) \right]_c,
\end{equation}
that depends solely on $|\tau|$ and $|\ell|$. 

Let us first restrict our attention to the purely spatial correlation function $\mathcal{C}_{\ln \varepsilon^\nu}(0,\ell)$. According to \eqref{eq:CorrLogYaglom}, the spatial covariance is expected to behave as a logarithm in the inertial range, up to an additive continuous and bounded function of the scale $f_\NS(\ell)$. In order to compare different DNSs at various Reynolds numbers, we superimpose on a master curve the logarithmic behaviors that will be eventually observed. Indeed, we have checked that, for the accessible scales of the inertial range $\eta_K\ll |\ell|\le L$ and for all Reynolds numbers, we can extract a modified integral length scale $L^\star_{\NS}$ that can be seen as the absorption of that remaining additive correction (data not shown). In other words, for this range of scales, $f_\NS(\ell)\approx f_\NS(0)$, such that $L^\star_{\NS}\approx L\exp(f_{\NS}(0))$.

The resulting spatial covariances are shown in Figure~\ref{fig:DNS}(c) as functions of $|\ell|/L^\star_{\NS}$ for the different Reynolds numbers, together with the reference law $\mu \ln(L^\star_{\NS}/|\ell|)$ plotted as a black dashed line with $\mu=0.21$. Several remarks are in order. First, as the Reynolds number increases, the range of scales over which the estimated covariance follows the logarithmic prediction becomes progressively wider, as it is expected from the formerly depicted phenomenology of fluid turbulence, and the Reynolds number dependence of the Kolmogorov dissipative length scale $\eta_K$ \eqref{eq:EtaK}. Second, at dissipative scales $|\ell|\le \eta_K$, the covariance grows faster than logarithmically. This is a usual phenomenon that is encountered in fluid turbulence when reaching the dissipative range, and can be fairly well understood in the following way. It is worth noting that, rather than exhibiting purely logarithmic behavior, numerical simulations in \cite{BecMuk25} have alternatively shown that the covariance $\mathcal{C}_{\ln \varepsilon^\nu}(0,\ell)$  scales as the square of the logarithm of the scale over a certain range. This observation still requires further understanding, interpretation, and confirmation at higher Reynolds numbers.

The dimensional arguments leading to the expression of the dissipative length scale proposed in \eqref{eq:EtaK} are mostly based on the initial phenomenology of Kolmogorov \cite{Kol41,Fri95} and need to be refined in presence of intermittency, as it is required by the following investigations of  Kolmogorov and Obukhov \cite{Kol62,Obu62}. An especially adapted language to deal with intermittent corrections has been developed in the context of the Multifractal formalism \cite{Fri95} and its generalization to take into account finite-Reynolds number corrections \cite{PalVul87,Nel90}. As a consequence of the intermittent (or multifractal) nature of turbulent fluctuations, the dissipative length scale $\eta_K$ \eqref{eq:EtaK} is no more unique and must also be considered as fluctuating in space. For a clear review on these aspects, we invite the reader to take a look at the article of Borgas \cite{Bor93}, and to Appendix~\ref{app:Laplace} where the main ingredients are recalled for convenience. As far as we are concerned, the fluctuating nature of the dissipative length scale has an important consequence on the variance of the logarithm of the dissipation field, which could be naively obtained using the covariance given \eqref{eq:CorrLogYaglom} at the scale $\ell=\eta_K$ yielding $\mathbb{V}\left[ \ln \varepsilon^\nu\right]\sim 3 \mu/4 \ln\mathcal R_e$. Doing so, we would get an under-estimation of the variance. Instead, a more realistic but more involved estimation is given by
 \begin{equation}\label{eq:EstimVarLogMF}
\mathbb{V}\left[ \ln \varepsilon^\nu\right]\build{\sim}_{\mathcal R_e\to\infty}^{}\dfrac{3 \mu}{4} \dfrac{(1+\mu/4)^2}{(1+\mu/8)^3}\ln\mathcal R_e.
 \end{equation}
 The derivation of \eqref{eq:EstimVarLogMF} is detailed in Appendix~\ref{app:Laplace}. Note that whenever $0 \le \mu \le 4(\sqrt{5}+1)$ the $\mu$ dependent pre-factor is larger than $3\mu/4$. The multifractal formalism therefore predicts a variance of $\ln \varepsilon^\nu$ larger than the one inferred from the inertial-range covariance alone. This is clearly observed in the present DNSs, as it is displayed in  Figure~\ref{fig:DNS}(c).

We finally turn to the Eulerian temporal correlation structure of $\ln \varepsilon^\nu$ while estimating the correlation function 
	$\mathcal{C}_{\ln \varepsilon^\nu}(\tau,0)$ \eqref{eq:DefCorrSTLogDiss} on the time line from the $N=1024^3$ DNS. In Figure~\ref{fig:DNS}(d), the temporal covariance of $\ln \varepsilon^\nu$ is shown in blue and compared with the spatial covariance shown in black. This comparison strongly suggests that the logarithmic covariance structure observed in space also extends to the temporal direction. In analogy with the spatial case, we therefore postulate that, in the inviscid limit,
\begin{align}\label{eq:CorrLogTemporal}
	\lim_{\nu\to 0} \mathcal{C}_{\ln \varepsilon^\nu}(\tau,0)= \mu \ln_+\left( \frac{T}{|\tau|}\right) + \mu g_{\NS}(\tau),
\end{align}
where $g_\NS$ is a bounded continuous function, the intermittency coefficient $\mu$ is the same as in the spatial covariance, and $T\simeq L/\sigma$. Analogously to the spatial case, we define an effective timescale $T^\star_\NS=T\exp(g_\NS(0))$ in the inviscid limit.
In Figure~\ref{fig:DNS}(d), the spatial covariance (black solid line) is represented as a function of $\xi=|\ell|/L^\star_\NS$, while the temporal one (blue solid line) is represented as a function of $\xi=\tau/T^\star_\NS$. The dashed black line indicates the logarithmic law $\mu\ln(1/\xi)$ with $\mu=0.21$. Although the behaviors at viscous and integral scales differ in space and in time, the inertial-range trends coincide. This observation will be the key guideline for the spatio-temporal stochastic modeling that we will be presenting in the next section. 

Let us notice that, instead of a logarithmic behavior \eqref{eq:CorrLogTemporal}, it was reported in \cite{PopChe90} an exponential behavior for the temporal covariance structure of the logarithm of dissipation and pseudo-dissipation. As discussed and developed in \cite{HuaSch14} in a slightly different context, it is argued that at these Reynolds numbers, it is barely possible to make a difference between a logarithmic and exponential functional dependence. Nonetheless, Figure~\ref{fig:DNS}(d) shows that the logarithmic trend is similar both in space and in time. As latterly argued in \cite{Che17,PerMor18,LetGou21,Zam22}, a logarithmic behavior is crucial to obtain a multifractal structure, whereas exponential trends cannot be extrapolated to the infinite Reynolds number limit.

The present analysis supports the statistical features of turbulent dissipation that have already been documented in the literature and recalled in Section \ref{Sec:StatStructDiss}. In particular, the one-point distribution of $\ln \varepsilon^\nu$ remain close to Gaussian, consistently with a lognormal description of $\varepsilon^\nu$, while the spatial covariance of $\ln \varepsilon^\nu$ is again found to display the expected logarithmic behavior \eqref{eq:CorrLogYaglom}, up to finite-Reynolds-number effects and statistical convergence. Extending this analysis to the time-resolved $1024^3$ DNS, we further observe that the Eulerian temporal covariance of $\ln \varepsilon^\nu$ is also compatible with a logarithmic behavior of the form \eqref{eq:CorrLogTemporal}, with the same intermittency coefficient $\mu$ as in the spatial case. To the best of our knowledge, this temporal counterpart has not been clearly evidenced previously.
As it will be developed in the next section, the purpose of the present work is to extend the arguments of the GMC theory to a spatio-temporal framework, able to give a meaning to such a  lognormal field, that follows a logarithmic correlation structure both in space \eqref{eq:CorrLogYaglom} and in time \eqref{eq:CorrLogTemporal}. In particular, in addition to proposing a construction method based on a limiting procedure, we propose a stochastic evolution able to reproduce the aforementioned statistical temporal structure, that will turn out to be realized by a Markovian evolution of the spatial Fourier modes.

\section{Introduction of the Gaussian Multiplicative Chaos}

\subsection{The GMC as a statistically homogeneous stochastic model of the dissipation field} \label{Sec:SpatialGMC}

As early understood by Yaglom \cite{Yag66}, the discrete multiplicative process, consisting in representing the dissipation field at a given position as the multiplication of positive random weights (i.e. multipliers) distributed over a dyadic structure would generate a covariance function behaving as a logarithm as it is proposed in \eqref{eq:CorrLogYaglom}.  Also early understood by him and soon after by Mandelbrot \cite{Man72,Man74,KahPey76}, such an iterative construction eventually generates a field which is not statistically homogeneous, meaning that its probability law is not invariant by translations. As a consequence, in a spatial modeling context, the respective correlation function defined in \eqref{eq:CorrLogYaglom} is not solely a function of the difference of the positions.

For these reasons, exploiting the additional assumption proposed in \cite{Kol62,Obu62} concerning the lognormal nature of the dissipation field, Mandelbrot proposed to define the dissipation field as being the exponential of a Gaussian field, which is required to be log-correlated. This model, that he called the \textit{limit lognormal model}, and that latterly Kahane \cite{Kah85} named the Gaussian Multiplicative Chaos (GMC), will turn out as we will see to be a statistically homogeneous stochastic representation of the aforementioned prescription of the dissipation field  \cite{Kol62,Obu62}. We invite the reader to consult early applications of the GMC for the modeling of rain and clouds \cite{SchLov87}, and the turbulent velocity field, both in an Eulerian \cite{SchVal92} and Lagrangian \cite{BacDel01,MorDel02,MorDel03,VigFri20} formulation.  Several mathematically inclined reviews have been since then devoted to this subject, including \cite{Ost04,RobVar10,RhoVar14,Sha16,Ber17} and to extensions beyond the lognormal framework \cite{SchMar01,BarMan02,MuzBac02,BacMuz03,ChaRie05}.

In simple words and in a formal manner, given a real number $\gamma\in\R$, a GMC $M_\gamma(x)$ with $x\in\R^d$ is the random field defined as the exponential of a zero-average log-correlated Gaussian field, i.e.
\begin{align}\label{eq:DefFormGMC}
M_\gamma(x) ``=" e^{\gamma X(x)}, 
\end{align}
where $X(x)$ is Gaussian, of zero-average and fully determined by its covariance function
\begin{align}\label{eq:DefFormCorrX}
\E \left[X(x) X(y)\right] ``=" \ln_+ \left(\frac{L}{|x-y|}\right).
\end{align}
Let us notice that the two equalities displayed in \eqref{eq:DefFormGMC} and \eqref{eq:DefFormCorrX} are put in quotes because they don't literally make sense. Indeed, such a Gaussian field $X$ \eqref{eq:DefFormCorrX}, if it exists, must be understood in a distributional manner since it is expected to be of infinite variance and thus cannot be considered pointwise. The meaning of exponentiating it  \eqref{eq:DefFormGMC} is even more questionable.  At this stage, the length scale $L$ has little importance, although it gives the dimension of a length to a position $x$ and will take the important meaning of the integral length scale when dealing with physical applications. The theory of the GMC consists in giving a proper meaning to the equalities \eqref{eq:DefFormGMC} and \eqref{eq:DefFormCorrX}, and in particular giving a range of possible values to the additional parameter $\gamma$ entering in the formulation. This will be done using a regularizing procedure, over a length scale $\epsilon$ and taking in an appropriate manner the limit $\epsilon\to 0$. It goes as follows.

Consider the random field $M_\gamma^\epsilon(x)$ for $x\in\R^d$ defined by
\begin{align}\label{eq:DefGammaEpsilon}
M_\gamma^\epsilon(x) = e^{\gamma X^\epsilon(x)-\frac{\gamma^2}{2}\E\left[\left( X^\epsilon\right)^2\right]}, 
\end{align}
with $X^\epsilon$ a given statistically homogeneous zero-average Gaussian field, thus fully determined by its covariance, assuming $\epsilon$ much smaller than any length scales, which is given by a positive definite function of the form
\begin{align}\label{eq:DefCorrXEpsilon}
\mathcal C_{X^\epsilon}(x-y)\equiv \E \left[X^\epsilon(x) X^\epsilon(y)\right] \approx  \ln_+ \left(\frac{L}{\epsilon+|x-y|}\right) + f_{X^\epsilon}(x-y),  
\end{align}
where $f_{X^\epsilon}$ is a bounded isotropic function of its argument, continuous at the origin, and furthermore bounded with $\epsilon$. As it is clear from \eqref{eq:DefCorrXEpsilon}, the variance of the field $X^\epsilon$ diverges with $\epsilon$ in a logarithmic fashion.  Let us remark that, for a finite $\epsilon>0$, if the Gaussian field $X^\epsilon$ exists, then its exponential and the respective field $M_\gamma^\epsilon$ \eqref{eq:DefGammaEpsilon} have a meaning in a classical sense. Moreover, by construction, $M_\gamma^\epsilon$ is of unit average for any values of the parameter $\gamma$ and for a finite $\epsilon>0$, i.e.
\begin{align}\label{eq:AverageMEpsilon}
\E M_\gamma^\epsilon(x) = 1.
\end{align}

The zero-average Gaussian field $X^\epsilon$ under consideration is fully characterized by its covariance $\mathcal C_{X^\epsilon}$  \eqref{eq:DefCorrXEpsilon} which is a function of solely the difference of the positions. To this regard, it is thus statistically homogeneous, and can be expressed as a Wiener integral according to 
 \begin{align}\label{eq:DefXEpsilonWiener} 
X^\epsilon(x) = \int_{\R^d}G_L^{\epsilon}(x-z) dW(z),
\end{align}
where $dW$ is a Gaussian white noise, i.e. the increment of the Wiener process over $dz$. The former is fully determined by its action on deterministic appropriate functions $f$ and $g$ such that
 \begin{align}\label{eq:DefdWAverage} 
\E\int_{\R^d}f(z)dW(z) = 0,
\end{align}
and
 \begin{align}\label{eq:DefdWCovariance} 
\E\left[\int_{\R^d}f(z)dW(z)\int_{\R^d}g(z)dW(z)\right] = \int_{\R^d}f(z)g(z)dz.
\end{align}
As a consequence of \eqref{eq:DefdWAverage} and \eqref{eq:DefdWCovariance}, the covariance of $X^\epsilon$ is given by
 \begin{align}\label{eq:CovXEpsilonFromWiener} 
\mathcal C_{X^\epsilon}(x-y) &= \int_{\R^d}G_L^{\epsilon}(z) G_L^{\epsilon}(x-y+z) dz\notag\\
&=\int_{\R^d}e^{2i\pi k\cdot (x-y)} \left|\widehat{G}_L^{\epsilon}(k) \right|^2dk,
\end{align}
where we have introduced the Fourier transform $\widehat{G}_L^{\epsilon}(k) $ of the kernel entering in \eqref{eq:DefXEpsilonWiener} and that reads
 \begin{align}\label{eq:FTGLEpsilon} 
\widehat{G}_L^{\epsilon}(k) =\int_{\R^d}e^{-2i\pi k\cdot x} G_L^{\epsilon}(x) dx.
\end{align}

We now need to choose a deterministic kernel $G_L^{\epsilon}(x) $ such that the implied covariance of $X^\epsilon$ coincides with the one depicted in \eqref{eq:DefCorrXEpsilon}. Although designing such a kernel could be done in the physical space, we find it more convenient to do it in the Fourier space \eqref{eq:FTGLEpsilon}. Two simple constraints on $\widehat{G}_L^{\epsilon}(k) $ that would lead to the behavior depicted in \eqref{eq:DefCorrXEpsilon} are the integrability of $|\widehat{G}_L^{\epsilon}|^2$ over $\R^d$ for any finite $\epsilon>0$, and when $\epsilon\to 0$, a power-law decay of the form $A_{d-1}^{-1}|k|^{-d}$ at large wave vector amplitude $|k|\to\infty$, with $A_{d-1} = 2\pi^{d/2}/\Gamma(d/2)$ the area of the boundary of the $d$-dimensional unit ball. This is a typical behavior required in the construction of fractional Gaussian fields \cite{Kra68,Kra70} of vanishing Hurst exponent. These Gaussian fields, characterized by a fractional regularity in a statistical sense, are of tremendous importance in the stochastic modeling of fluid turbulence, and can be seen as high dimensional statistically homogeneous versions of fractional Brownian motions \cite{Kol40,ManVan68}. The case of vanishing Hurst exponent is fully treated in \cite{DupRho17}. In this context, let us also mention that the GMC can also be defined as the limit of vanishing Hurst exponent of the exponential of a fractional Gaussian field \cite{HagNeu22}.

To set ideas, we propose to derive the statistical properties of the implied Gaussian field $X^\epsilon$ when assuming the simple situation of the following real, in particular Hermitian symmetric, function 
 \begin{align}\label{eq:ExampleFTGLEpsilonText} 
\widehat{G}_L^{\epsilon}(k) =\frac{1}{\sqrt{A_{d-1}}}\frac{1}{|k|^{d/2}}\mathds{1}_{1/L\le |k|\le 1/\epsilon},
\end{align}
where it is understood that $\epsilon<L$.  We show in Appendix \ref{App:CorrLog} that such a kernel $\widehat{G}_L^{\epsilon}(k)$ \eqref{eq:ExampleFTGLEpsilonText} leads to a logarithmic divergence of the variance, i.e.
 \begin{align}\label{eq:VarXEpsilonFromExKernel} 
\E\left[(X^\epsilon)^2\right]= \ln\frac{L}{\epsilon},
\end{align}
which furthermore says that the value at the origin of the bounded function entering  in \eqref{eq:DefCorrXEpsilon}  vanishes, ie. $f_{X^\epsilon}(0) =0$. Although the variance diverges with $\epsilon$, the covariance will remain bounded away from the origin for any $\epsilon$, and we obtain for a given space lag $|\ell|>0$ that the expected logarithmic behavior
\begin{align}\label{eq:ResCXEll}
\mathcal C_{X}(\ell)=\lim_{\epsilon\to 0}\E \left[X^\epsilon(x) X^\epsilon(x+\ell)\right]=\ln_+\left(\frac{L}{|\ell|} \right)+ f_X(\ell),
\end{align}
where $f_X(\ell)=f_X(|\ell|)$ is a continuous and bounded function of its arguments. We believe that using a smooth in $k$ cut-off function at the wave vector amplitude $L^{-1}$ and $\epsilon^{-1}$ instead of the sharp one as it is chosen in \eqref{eq:ExampleFTGLEpsilonText} would not change the global behavior depicted in \eqref{eq:DefCorrXEpsilon}, in particular the logarithmic behavior will remain unchanged. Although, it would have some consequences on the very expression of the additional bounded function $f_{X^\epsilon}$ entering in the RHS of \eqref{eq:DefCorrXEpsilon}, and of its limiting behavior $f_{X}$ as $\epsilon\to 0$ that enters in \eqref{eq:ResCXEll}. For the sake of completeness, we compute in Appendix~\ref{App:CorrLog} the expression of this additional continuous and bounded function $f_X(\ell)$ that enters in \eqref{eq:ResCXEll} for the particular example of the kernel $\widehat{G}_L^{\epsilon}$ proposed in \eqref{eq:ExampleFTGLEpsilonText}. In particular, we show that it is continuous in the vicinity of the origin, its precise value being dependent on space dimension \eqref{eq:ValueFXAtOrigin}.

 Following this regularizing procedure allows to give a meaning to the formal equality provided in \eqref{eq:DefFormGMC}. As it is reviewed in \cite{Ost04,RobVar10,RhoVar14,Sha16,Ber17}, the GMC is the positive random measure $M_\gamma(x)$ defined as the limit as $\epsilon\to 0$ of the regularized measure $M_\gamma^\epsilon(x)$ \eqref{eq:DefGammaEpsilon}, i.e.
\begin{align}\label{eq:DefGMCasLimMepsilon}
M_\gamma(x) = \lim_{\epsilon\to 0} e^{\gamma X^\epsilon(x)-\frac{\gamma^2}{2}\E\left[\left( X^\epsilon\right)^2\right]}.
\end{align}
It is shown in the seminal work of Kahane \cite{Kah85} that the measure is non-degenerate (i.e. does not reduce to zero in a loose sense) only when $\gamma^2<2d$. For this range of values of $\gamma$, the GMC is of unit-average, and the equality displayed in \eqref{eq:AverageMEpsilon} remains true in the limit. Once again, we invite the reader to more mathematically inclined articles for a precise meaning of degeneracy and a more proper definition of the GMC than \eqref{eq:DefGMCasLimMepsilon},  that has to be integrated against a test function in order to deal with its distributional nature \cite{Ost04,RobVar10,RhoVar14,Sha16,Ber17}.

Going back to the physics of turbulence, the GMC $M_\gamma$ \eqref{eq:DefGMCasLimMepsilon}, and its regularized version 
$M_\gamma^\epsilon$ \eqref{eq:DefGammaEpsilon}, are especially well adapted to reproduce the statistical behaviors of respectively the dissipation field at infinite Reynolds number, and its counterpart $\varepsilon^\nu$ \eqref{eq:DefDissField} at a finite viscosity, using the correspondence, for a given average dissipation $\overline{\varepsilon}$ \eqref{eq:FinitAverageDissField},
\begin{align}
\varepsilon^\nu(.,x) &= \overline{\varepsilon}M_\gamma^\epsilon(x),\label{eq:CorrModDissMepsilon}\\
\mu&=\gamma^2,\label{eq:CorrMuGamma2}\\
\eta_K(\nu)&=\epsilon \label{eq:CorrEtaKEpsilon},
\end{align}
where the Kolmogorov dissipative length scale is defined in \eqref{eq:EtaK}. In particular, in the limit $\eta_K=\epsilon\to 0$ (or equivalently $\nu\to 0$), we get that the moments of the locally averaged dissipation field $\varepsilon^\nu_\ell $ \eqref{eq:DefDissFieldOverBall} behave as a power-law \eqref{eq:PowerLawMoments}, with a quadratic spectrum $\tau_q$ \eqref{eq:LNTAUQ} and same multiplicative constant $c_q$ \eqref{eq:ExpCqLN}, which is finite for the same sharp condition \eqref{eq:SharpCondCq}. Let us also mention that, interestingly, the GMC \eqref{eq:DefGMCasLimMepsilon} can also be considered in Fourier space. Its rigorous harmonic analysis is proposed in \cite{GarVar26}.

\subsection{The spatio-temporal GMC} \label{Sec:SpatioTempGMC}

The modeling correspondence provided in \eqref{eq:CorrModDissMepsilon} is ambiguous because the temporal dependence of the dissipation field $\varepsilon^\nu$ has been left aside. For this reason, we would like now to make a proposition for the stochastic modeling of the dissipation based on a spatio-temporal generalization of the GMC.

A natural way to do this is to consider a spatio-temporal extension $X^\epsilon(t,x)$ of the regularized field entering in the definition of the GMC \eqref{eq:DefGMCasLimMepsilon}. To do so, we will be using a related construction that has been developed in \cite{ChaGaw03} (see also earlier propositions in \cite{HolSig94,KomPap97,FanKom00}, and related discussions in  \cite{KomPes04,EyiBen13}), recently revisited in \cite{ChaLem26}.

Let us introduce the Gaussian space-time white noise $dW(t,x)$ which is determined in a similar way as in \eqref{eq:DefdWAverage} and \eqref{eq:DefdWCovariance}, but with an additional integration over time. We obtain, for any appropriate deterministic test functions $f(t,x)$ and $g(t,x)$,
\begin{align}\label{eq:DefdWSTAverage} 
\E\int_{\R\times\R^{d}}f(t,x)dW(t,x) = 0,
\end{align}
and
 \begin{align}\label{eq:DefdWSTCovariance} 
\E\left[\int_{\R\times\R^{d}}f(t,x)dW(t,x) \int_{\R\times\R^{d}}g(t,x)dW(t,x) \right] = \int_{\R\times\R^{d}}f(t,x)g(t,x)dtdx.
\end{align}
As it will become clear later, we will need to define the Fourier transform $\widehat{f}(t,k)$ of any appropriate functions $f(t,x)$, similarly to \eqref{eq:FTGLEpsilon},  as 
\begin{align}\label{eq:DefFTR} 
\widehat{f}(t,k) = \int_{\R^d}e^{-2i\pi k\cdot x}f(t,x)dx.
\end{align}
When we will be considering statistically homogeneous fields, which are not in particular square-integrable, the respective Fourier transform will have to be considered in a distributional manner in the $k$-variable. Furthermore, the Fourier transform $\widehat{dW}(t,k)$ of the space-time white noise $dW(t,x)$ should also be considered in a distributional manner, both in the wave vector $k$-variable, but also in time. As it is recalled in \cite{ChaLem26}, when fields are formulated on the torus $\T^d$ (i.e. in periodic space), respective Fourier modes are of finite variance, the remaining distributional nature in time of  $\widehat{dW}(t,k)$ will be considered in an appropriate manner when dealing with stochastic evolutions. In the following, we will keep in mind these aspects and nonetheless take the short-cut of considering the continuous Fourier transform for the sake of simplicity.

Consider then the spatio-temporal random field $X_\beta^\epsilon(t,x)$ which Fourier transform $\widehat{X_\beta^\epsilon}(t,k)$ \eqref{eq:DefFTR}  evolves according to the Markovian dynamics of Ornstein-Uhlenbeck type
\begin{align}\label{eq:OUFTX}
d\widehat{X_\beta^\epsilon}(t,k) = -\frac{1}{T_{k,\beta}}\widehat{X_\beta^\epsilon}(t,k) dt +\widehat{G}_L^{\epsilon}(k) \sqrt{\frac{2}{T_{k,\beta}}}\widehat{dW}(t,k),
\end{align}
for a given initial condition $\widehat{X_\beta^\epsilon}(0,k)$. We use in \eqref{eq:OUFTX} the same Fourier transform $\widehat{G}_L^{\epsilon}(k) $ of the deterministic kernel  entering in the construction of the spatial GMC \eqref{eq:FTGLEpsilon}  (take for instance the proposition made in \eqref{eq:ExampleFTGLEpsilonText}), and a $k$-dependent correlation time scale $T_{k,\beta}$ asked to be a decreasing function of the wave vector amplitude $|k|$. The former reads, following the notation of \cite{ChaGaw03} (see also the devoted discussion in \cite{ChaLem26}),
\begin{align}\label{eq:DefTk}
T_{k,\beta}= \frac{1}{D_3 |k|^{2\beta}},
\end{align}
where $\beta>0$ is a free parameter of the formulation, and $D_3>0$ a constant that has the dimension of  time$^{-1}$ $\times$ length$^{2\beta}$. The positivity of $\beta$ ensures that the Fourier transform at large wave numbers is correlated on a shorter time scale than the ones at lower wave numbers.

Starting for instance from the vanishing initial condition $X_\beta^\epsilon(0,x)=0$, or equivalently $\widehat{X_\beta^\epsilon}(0,k)=0$, the evolving system \eqref{eq:OUFTX} eventually reaches a statistically homogeneous and stationary regime in which the process $X_\beta^\epsilon(t,x)$ is a zero-average Gaussian process, which Fourier transform reads
\begin{align}\label{eq:StatStatioSolOUFTX}
\widehat{X_\beta^\epsilon}(t,k) = \widehat{G}_L^{\epsilon}(k) \sqrt{\frac{2}{T_{k,\beta}}}\int_{-\infty}^t e^{-\frac{t-s}{T_{k,\beta}}}\widehat{dW}(s,k).
\end{align}
In this regime, the random field $X_\beta^\epsilon$ is of zero-average and is characterized by the covariance
\begin{align}\label{eq:CovXEpsilonST} 
\mathcal C_{X_\beta^\epsilon}(\tau,\ell) &= \E\left[X_\beta^\epsilon(t,x)X_\beta^\epsilon(t+\tau,x+\ell) \right]\notag \\
&=\int_{\R^d}e^{2i\pi k\cdot \ell}e^{-\frac{|\tau|}{T_{k,\beta}}}\left|\widehat{G}_L^{\epsilon}(k)\right|^2 dk,
\end{align}
which is obtained while injecting the expression \eqref{eq:StatStatioSolOUFTX} and taking the expectation using the covariance structure of the Fourier transform of the white noise.  At a vanishing time lag $\tau=0$, the covariance function $\mathcal C_{X_\beta^\epsilon}(0,\ell)$ \eqref{eq:CovXEpsilonST} coincides with the covariance of the formerly defined field \eqref{eq:CovXEpsilonFromWiener}, independently of the value of the parameter $\beta$ and with the logarithmic structure depicted in \eqref{eq:DefCorrXEpsilon}. In the limit $\epsilon\to 0$, for $|\tau|$ and $|\ell|$ strictly positive, the spatio-temporal covariance becomes
\begin{align}\label{eq:CXBetaAsympt}
\mathcal C_{X_\beta}(\tau,\ell)=\lim_{\epsilon\to 0} \mathcal C_{X_\beta^\epsilon}(\tau,\ell) &=\int_{\R^d}e^{2i\pi k\cdot \ell}e^{-\frac{|\tau|}{T_{k,\beta}}}\left|\widehat{G}_L(k)\right|^2 dk.
\end{align}
We show in Appendix \ref{App:CorrLogST}, for the particular choice made in \eqref{eq:ExampleFTGLEpsilonText}, that
\begin{align}\label{eq:ResCXTauEllMax}
\mathcal C_{X_\beta}(\tau,\ell)=\ln_+\left(\frac{L}{\max\left[2\pi |\ell|,(D_3|\tau|)^{\frac{1}{2\beta}}\right]} \right)+ f_{X_\beta}(\tau,\ell),
\end{align}
where $f_{X_\beta}(\tau,\ell)=f_{X_\beta}(|\tau|,|\ell|)$ is a bounded function of its arguments.

Notice first that at vanishing time lags $\tau=0$, the additional bounded function entering in \eqref{eq:ResCXTauEllMax} coincides, up to an additive constant, with its purely spatial counterpart \eqref{eq:ResCXEll}, i.e. $f_{X_\beta}(0,\ell)=f_{X}(\ell)+\ln(2\pi)$, independently of $\beta$. As developed in \eqref{eq:TimeLineST}, along the time line we obtain
\begin{align}\label{eq:TimeLineSTtext} 
\mathcal C_{X_\beta}(\tau,0) &= \frac{1}{2\beta}\ln_+\left(\frac{L^{2\beta}}{D_3|\tau|} \right)+f_{X_\beta}(\tau,0).
\end{align}
Interestingly, whereas it can be shown that the function $f_{X_\beta}(|\tau|,0)$ of the single variable $|\tau|$ is continuous, the function $f_{X_\beta}(|\tau|,|\ell|)$ of the two variables $|\tau|$ and $|\ell|$ is not \eqref{eq:FXSpaceFXTempNotEqual}. In other words, when comparing the expressions provided in \eqref{eq:DefFXTemp} and  \eqref{eq:DefFXSpace}, we have 
\begin{align}\label{eq:DiscontFTauEll} 
f_{X_\beta}^{\text{\tiny{temp}}} \equiv \lim_{|\tau|\to 0}f_{X_\beta}(|\tau|,0) \neq \lim_{|\ell|\to 0}f_{X_\beta}(0,|\ell|)\equiv f_{X_\beta}^{\text{\tiny{space}}},
\end{align}
for any space dimension $d$ and any $\beta>0$.

The spatio-temporal GMC $M_{\gamma,\beta}(t,x)$ is readily obtained as the limit of the exponential of the logarithmically correlated Gaussian field $X_\beta^\epsilon(t,x)$, that reads
\begin{align}\label{eq:STGMC}
M_{\gamma,\beta}(t,x) = \lim_{\epsilon\to 0}M_{\gamma,\beta}^\epsilon(t,x), 
\end{align}
where
\begin{align}\label{eq:STGMCEpsilon}
M_{\gamma,\beta}^\epsilon(t,x) =e^{\gamma X_\beta^\epsilon(t,x)-\frac{\gamma^2}{2}\E \big[(X_\beta^\epsilon)^2\big]}.
\end{align}

Going back to the stochastic modeling of turbulence, and in particular of the spatio-temporal structure of the dissipation field \eqref{eq:DefCorrSTLogDiss}, observations made on DNSs of the Navier-Stokes in Section \ref{Sec:NumObservationsHopkins} suggest that the logarithm of dissipation field is correlated both in space and time in a logarithmic way with the spatial lag $|\ell|$ \eqref{eq:CorrLogYaglom} and/or with the time lag $\tau$ \eqref{eq:CorrLogTemporal}, with the same proportionality constant $\mu$. This sets the free parameter $\beta$  to the value
\begin{align}\label{eq:SetBeta12}
\beta=\frac{1}{2},
\end{align}
in a similar way as it was concluded in \cite{ChaLem26} when proposing a similar time-evolving stochastic model of the turbulent velocity field, and justified by the necessity to introduce the sweeping effect in this random picture.

In this context, using the value of $\beta=\sfrac{1}{2}$ proposed in \eqref{eq:SetBeta12}, a generalization of the correspondence between the dissipation field $\varepsilon^\nu$ \eqref{eq:DefDissField}  and the standard GMC, as it is proposed in \eqref{eq:CorrModDissMepsilon}, \eqref{eq:CorrMuGamma2} and  \eqref{eq:CorrEtaKEpsilon}, would now use the proposed extension of the GMC made in \eqref{eq:STGMC} and \eqref{eq:STGMCEpsilon}, and would read in the present spatio-temporal framework as 
\begin{align}
\varepsilon^\nu(t,x) &= \overline{\varepsilon}M_{\gamma,\beta}^\epsilon(t,x),\label{eq:ModDissMepsilonST}\\
\mu&=\gamma^2,\label{eq:CorrMuGamma2ST}\\
\eta_K(\nu)&=\epsilon, \label{eq:CorrEtaKEpsilonST}\\
\frac{1}{2}&=\beta.\label{eq:SetBeta12List}
\end{align}
Notice that the correspondences concerning the intermittency coefficient \eqref{eq:CorrMuGamma2ST} and the regularizing scale \eqref{eq:CorrEtaKEpsilonST} remain unchanged compared to \eqref{eq:CorrMuGamma2} and \eqref{eq:CorrEtaKEpsilon}. In particular, in the limit $\eta_K=\epsilon\to 0$ (or equivalently $\nu\to 0$), we get that the moments of the locally averaged dissipation field $\varepsilon^\nu_\ell $ \eqref{eq:DefDissFieldOverBall} behave as a power-law \eqref{eq:PowerLawMoments}, with a quadratic spectrum $\tau_q$ \eqref{eq:LNTAUQ} and same multiplicative constant $c_q$ \eqref{eq:ExpCqLN}, which is finite for the same sharp condition given in \eqref{eq:SharpCondCq}.

\subsection{Simulations of the spatio-temporal GMC, and a final comparison against DNSs}
\label{Sec:SimSTGMCandDNS}

Let us now explain how numerical instances of the dynamics \eqref{eq:OUFTX} could be generated. For obvious computational purposes, we restrict ourselves to the one-dimensional setting $d=1$ and, consistently with the DNS observations of Section~\ref{Sec:NumObservationsHopkins}, we set the parameter $\beta$ to the value $\beta=1/2$ \eqref{eq:SetBeta12}, such that the logarithm of dissipation field is correlated both in space and time in a logarithmic way, with the spatial lag $|\ell|$ \eqref{eq:CorrLogYaglom} and/or with the time lag $\tau$ \eqref{eq:CorrLogTemporal}, with the same proportionality constant $\mu$.

In order to make use of the Discrete Fourier Transform (DFT), we need to formulate the model on the one-dimensional torus of length $\Lt$, $\mathbb{T}_{\Lt}=\mathbb{R}/(\Lt\mathbb{Z})$. This amounts to write the model on the segment $[0, \Lt]$ with periodic boundary conditions. This periodic version should be understood as an approximation of the model on the whole line, recovered in the asymptotic limit $\Lt\to\infty$. Although the model is considered here in a periodic setting, we keep the same notations as in the non-periodic case for the various fields and observables of interest. The Fourier modes are now indexed by discrete wave numbers $k\in \mathbb{Z}/\Lt$ and are driven by the Fourier modes of a space-periodic Gaussian white noise. In Appendix~\ref{App:PeriodicModel}, we specify the meaning of the model in a periodic setting and derive the asymptotic behaviors of the spatial and temporal correlation functions in this periodic framework and compare them to those of the model on $\mathbb{R}$. As discussed in Section~\ref{Sec:NumObservationsHopkins}, the behavior at the origin of $f_\NS(\ell)$ and $g_\NS(\tau)$ naturally defines a characteristic lengthscale and timescale. In the present periodic setting, the corresponding quantities are computed in Appendix~\ref{App:PeriodicModel} and denoted by $L_{X_\beta}^\star$ \eqref{eq:LstarPeriodic} and $T_{X_\beta}^\star$ \eqref{eq:TstarPeriodic}. These spatial and temporal scales depend explicitly on the domain length $\Lt$, but converge in the infinite domain size limit to the values expected for the model on $\mathbb{R}$, namely $L_{X_\beta}^\star \to \sfrac{Le^{-\varrho}}{2\pi}$ and $T_{X_\beta}^\star \to \sfrac{Le^{-\varrho}}{D_3}$ as $\Lt$ tends to infinity, and $\varrho$ is the Euler-Mascheroni constant.
The periodic formulation thus provides a natural and convenient approximation of the model on the real line.

For numerical simulations, the torus is further discretized with $N_x$ collocation points which allows one to use the discrete Fourier transform and evolve the Fourier modes numerically. Time integration is performed using the exact-in-law numerical scheme adapted from \cite{ChaLem26} and presented in Appendix~\ref{App:NumSchem}.

\begin{figure}[t]
	\centering
	\includegraphics[width=0.85\linewidth]{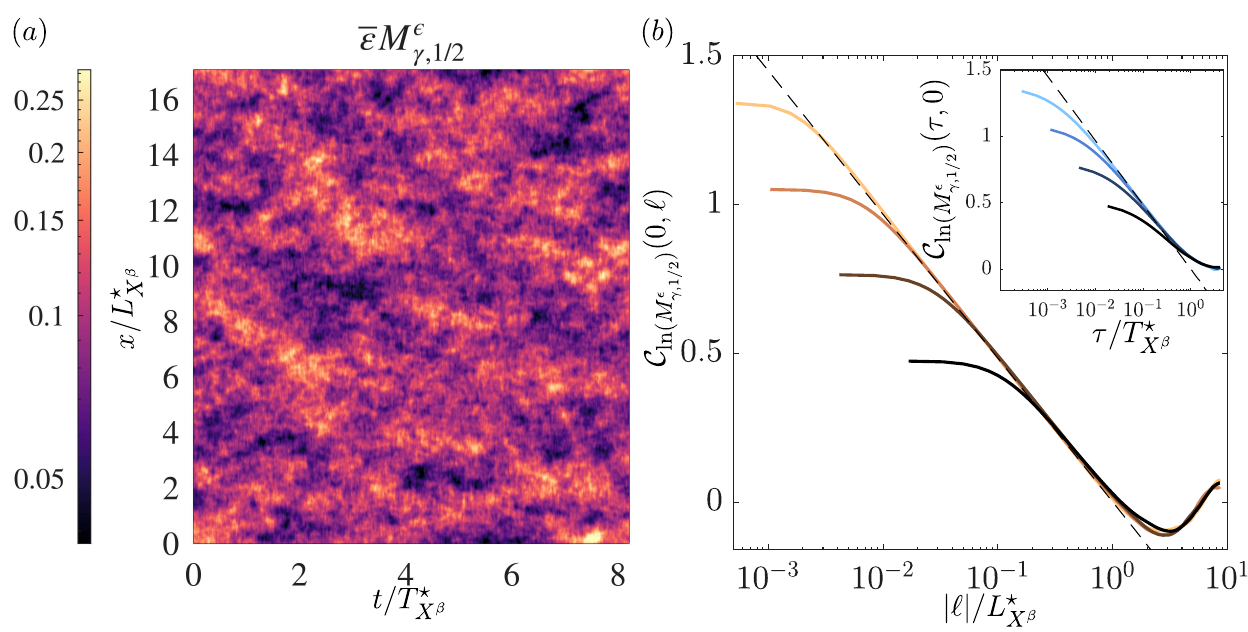}
	\caption{(a) Spatio-temporal diagram representing the evolution of $\overline{\varepsilon}M_{\gamma,1/2}^\epsilon$ as a function of time. The average dissipation rate $\overline{\varepsilon}$ is the same as the $1024^3$ DNS and other parameters, specified in the main text, are also chosen to match the $1024^3$ DNS. Additionally, we took $\epsilon= L^{\star}_{X_\beta}/10 $ and $N_x=1024$. The field displays large scale correlations in both space and time, as expected from the space-time logarithmic correlations. (b) Main figure displays spatial correlations for varying regularization parameter $\epsilon=L^{\star}_{X_\beta}/10$ (black solid line), $L^{\star}_{X_\beta}/40$, $L^{\star}_{X_\beta}/160$ and $\epsilon=L^{\star}_{X_\beta}/640$ (yellow solid line) versus the normalized spatial lag $|\ell|/ L_{X_\beta}^\star$. In the inset we display the corresponding time correlations as a function of $\tau/T^{\star}_{X_\beta}$ for $\epsilon$ varying from $\epsilon=L^{\star}_{X_\beta}/10$ (black solid line) down to $\epsilon=L^{\star}_{X_\beta}/640$ (light blue solid line).  $ L_{X_\beta}^\star$ and $ T_{X_\beta}^\star$ are respectively defined by \eqref{eq:LstarPeriodic} and \eqref{eq:TstarPeriodic} and the number of collocation points are taken as $N_x=2^{10}, \, 2^{12}, 2^{14} $ and $2^{16}$ for the smallest value of $\epsilon$. In the main figure and inset, the solid black curves are obtained with the same set of parameters as panel (a) mimicking the $1024^3$ DNS. In each figure, we superimpose in dashed black lines the expected logarithmic behavior of the correlations. }\label{fig:Model}
\end{figure}

At this stage, the free parameters of the model must be specified. First, $\gamma$, $L$ and $D_3$ are chosen so as to match the inertial-range behavior of the $1024^3$ DNS. The parameter $\gamma$ is naturally set to $\sqrt{\mu}$ in order to match the overall prefactor of the logarithmic correlations \eqref{eq:CorrLogYaglom}. Next, to ensure that the inertial ranges of the model coincide with those inferred from the DNS, we impose
$L=L_\NS^\star\exp(-f_{X_\beta}^{ \text{\tiny space }})$, and $D_3=(\sfrac{L}{T_\NS^\star})\exp(f_{X_\beta}^{ \text{\tiny temp }})$, where $L_\NS^\star$ and $T_\NS^\star$ have been estimated in Section~\ref{Sec:NumObservationsHopkins} and the expressions of $f_{X_\beta}^{ \text{\tiny temp }}$ and $f_{X_\beta}^{ \text{\tiny space }}$ are derived in Appendix~\ref{App:CorrLogST}. With these choices, one indeed recovers matching inertial ranges between the $1024^3$ DNS and the model in virtue of \eqref{eq:CorrLogTemporal}, \eqref{eq:TimeLineSTtext} for time and \eqref{eq:CorrLogYaglom} and \eqref{eq:ResCXEll} in space. Finally, the values used for $\epsilon$ and the numbers of collocation points $N_x$ are given in the caption of Figure~\ref{fig:Model}.

 In every simulation, we choose $\Lt=5L_{X_\beta}^\star$ and a time step $\delta t=T_{k_{\mathrm{max}},\beta}/5$, where $k_{\mathrm{max}}=N_x/2$ is the largest resolved Fourier mode. Finally, in order to avoid Gibbs phenomenon caused by a discontinuity at large wave numbers in Fourier space, we use the Fourier kernel $
\widehat{G}_{L}^\epsilon(k)=\mathds{1}_{|k|\geq 1/L}e^{-\epsilon |k|}/\sqrt{2|k|}$.
In the vanishing-$\epsilon$ limit, the statistics do not depend on the precise shape of the small-scale regularization allowing us some freedom in the choice of the small scale regularization.
The initial condition of the dynamics is chosen in accordance with the invariant measure of the dynamics. By doing so, the statistically stationary regime is reached already at $t=0$ and we can perform averages without waiting for the end of a transient regime. The time integration is carried out over $200T^\star_{X_\beta}$ units of time. Spatial statistics are computed as time averages over the whole integration time and time statistics are computed on $25$ slices of $8 T^\star_{X_\beta}$ units of time.

A typical numerical realization obtained in this way is shown in Figure~\ref{fig:Model}(a). Panel (a) displays a spatio-temporal rendering of the model dissipation field $\overline{\varepsilon}\, M^\epsilon_{\gamma,1/2}(t,x)$ as a function of suitably rescaled time and space variables. The characteristic length and time are those defined by \eqref{eq:LstarPeriodic} and \eqref{eq:TstarPeriodic}. The average dissipation rate $\overline{\varepsilon}$ is estimated from the $1024^3$ DNS data. Several qualitative remarks can be made. First, as for the DNS field shown in Figure~\ref{fig:DNS}, the model exhibits correlations over large scales in both space and time, in agreement with the underlying logarithmic correlation structure. Second, in contrast with the DNS, these structures tend to appear and disappear within a given spatial region, which gives the overall picture a more patchy aspect than the filamentary behavior visible in Figure~\ref{fig:DNS}. The present model does not incorporate the so-called sweeping effect, which can be interpreted as the advection of the smaller scales of the flow by the larger ones. Capturing this mechanism would require a dynamical coupling between Fourier modes, which is absent from the present construction. In the context of modeling the spatio-temporal structure of the velocity field itself \cite{ChaLem26}, rather than the dissipation field, the choice $\beta=1/2$ is imposed in order to reproduce sweeping at the statistical level. This choice is motivated there by evidence from DNS data and renormalization-group arguments of \cite{GorBal21}. The intermittency coefficient $\mu$ being the same for temporal and spatial correlations as exposed in Section~\ref{Sec:NumObservationsHopkins} can therefore be seen as a consequence of sweeping effect and also imposes $\beta=\sfrac 12$ in the model. 

Finally, one may note that the model field appears qualitatively rougher in time than in the DNS, which is expected to be smooth in time at finite viscosity. By contrast, for a finite value of the regularization parameter $\epsilon$, the field $X^\epsilon_\beta$ has in time the same regularity as a Brownian motion. As shown in \cite{ChaLem26}, this drawback can be overcome by introducing a carefully chosen forcing term that is smooth in time. 

Panel (b) shows the corresponding spatial correlation function of $\ln\!\big(\overline{\varepsilon} M^\epsilon_{\gamma,1/2}\big)$, which coincides with the covariance of $ \gamma X_\beta^\epsilon$, for different values of $\epsilon$, together with the associated temporal covariance shown in the inset. These quantities are represented in terms of the normalized variables $|\ell|/L^\star_{X_\beta}$ and $\tau/T^\star_{X_\beta}$, respectively. As in the DNS, both display the same logarithmic trend throughout the inertial range, with slope fixed by the intermittency coefficient $\mu$. This final comparison shows that the present spatio-temporal GMC construction reproduces, at the level of second-order statistics, the main spatial and temporal signatures observed in the Navier-Stokes data. With our choices of parameters, the model matches the DNS data in the inertial range. The large and small scales behavior are however not reproduced by the current model. The large scale behavior is set by the precise shape of the forcing of the DNS field and is therefore not universal. For the small scales however, the discrepancy between the model and DNS observations is due to viscous range intermittency discussed through the behavior of the logarithm of dissipation \eqref{eq:EstimVarLogMF} which is not accounted for by the model.

\section{Conclusion and perspectives}\label{Sec:Conclusion}

We have proposed a generalization of the GMC to a spatio-temporal framework able to reproduce the statistical structure of the turbulent dissipation field, as it is observed in a publicly available turbulence database. The present generalization is based on a Markovian stochastic dynamics, amenable to both a theoretical study and a numerical implementation. It is in particular consistent to the observed logarithmic structure in the inertial range of scales, both in space and in time, of the logarithm of the dissipation field.

As mentioned in the presentation of the simulations of the Navier-Stokes equations, the covariance structure of the dissipation field behaves in a non-trivial way in the dissipative range of scales. Although several aspects of the turbulent fluctuations in this viscous-dependent range of scales are correctly taken into account by the multifractal formalism, building a random field able to reproduce them is still fully opened. 

Such a spatio-temporal GMC will eventually be at the core of a forthcoming complete synthetic model of the turbulent velocity field, which underlying Gaussian structure has been proposed in \cite{ChaLem26}, and could help the training of data-driven \cite{WitGab26} and diffusion \cite{LiBuz26} models. We keep these possible improvements and developments for future investigations.

\vspace{1cm}

\textbf{Acknowledgments and fundings: }We thank Jean-Yves Chemin for his help concerning the Fourier transform of homogeneous functions of critical order, and Charles-Edouard Br\'ehier for discussions. L.C. is partially supported by the Simons Foundation Award ID: 651475, and by Agence Nationale de la Recherche ANR-20-CE30-0035.

\appendix

\section{Asymptotic behavior of the variance of $\ln \varepsilon^\nu$ using the multifractal formalism} \label{app:Laplace}

The multifractal formalism (MF) is a precise language that has been developed over the years in order to deal with the peculiar nature of the fluctuations encountered in fluid turbulence  \cite{Fri95}. We will be here following in particular the notation used in \cite{Bor93} that are especially well adapted when dealing with the fluctuation of the dissipation field $\varepsilon^\nu$, and more precisely on its coarse-grained version $\varepsilon^\nu_\ell$  \eqref{eq:DefDissFieldOverBall}. The purpose of this Appendix is the presentation of key ingredients of the MF. Doing so, we will be able to compute the variance of $\ln \varepsilon^\nu$ as the limit at vanishing scale $|\ell|\to 0$ of the corresponding variance $\ln \varepsilon^\nu_\ell$ which is eventually predicted by the MF. Note that the following probabilistic picture can be recasted into a random one using the developments of \cite{WarBen25}.

Accordingly, in order to interpret in a probabilistic manner the  behavior \eqref{eq:FinitHOMomentsDissField} of the moments of the coarse-grained dissipation field $\varepsilon^\nu_\ell$  \eqref{eq:DefDissFieldOverBall} in the inertial range, which corresponds to neglecting finite viscosity effects, it was early proposed the following equality \textit{in law}, meaning a equality relating two random variables that have same probability distributions,
\begin{align}\label{eq:ProbCGDissInertial}
	\varepsilon^\nu_\ell =\overline{\varepsilon}\left(\frac{\ell}{L}\right)^{\alpha-1}
\end{align}
where $\overline{\varepsilon}$ is the (viscosity independent) average dissipation \eqref{eq:FinitAverageDissField}, $L$ the integral length scale introduced through the forcing term in the Navier-Stokes equations \eqref{eq:NSforced}, and $\alpha$ a fluctuating exponent, assumed to fluctuate around an average value close to unity and over the range $[\alpha_{\min},\alpha_{\max}]$, which probability density function $\mathcal F_\ell (\alpha)$ reads
\begin{align}\label{eq:ProbAlphaCGDissInertial}
	\mathcal F_\ell (\alpha)=\frac{1}{\mathcal Z(\ell)}\left(\frac{\ell}{L}\right)^{1-f(\alpha)},
\end{align}
where $f(\alpha)$ is a viscosity-independent parameter function, known in the literature as the multifractal spectrum, and $\mathcal Z(\ell)$ a normalization ensuring that the density $\mathcal F_\ell$ is of unit integral at each scale $\ell$. A steepest-descent argument allows to relate the spectrum of exponents $\tau_q$  \eqref{eq:FinitHOMomentsDissField} and the multifractal spectrum $f(\alpha)$ through a Legendre transform that reads
\begin{align}
	\E \left[ \left(\varepsilon^\nu_\ell\right)^q \right] &= \frac{\overline{\varepsilon}^q}{\mathcal Z(\ell)}\int_{\alpha_{\min}}^{\alpha_{\max}}\left(\frac{\ell}{L}\right)^{q(\alpha-1)+1-f(\alpha)}d\alpha\\
	&\build{\sim}_{\ell\to 0}^{} c_q \overline{\varepsilon}^q \left(\frac{\ell}{L}\right)^{\tau_q},
\end{align}
with
\begin{align}\label{eq:Legendre}
	\tau_q = \min_{\alpha} \left[ q(\alpha-1)+1-f(\alpha)\right],
\end{align}
up to a multiplicative factor $c_q$ that can be computed while following the steepest-descent procedure. At this stage, the present formalism is quite general, and mostly depends on the precise shape of the multifractal spectrum $f(\alpha)$. In the particular situation where we assume a quadratic shape, i.e.
\begin{align}\label{eq:FAlphaLN}
	f(\alpha)=1-\frac{(\alpha-1-\mu/2)^2}{2\mu},
\end{align}
we recover the initial proposition of a quadratic spectrum of exponents  \eqref{eq:LNTAUQ} by  Kolmogorov and Obukhov \cite{Kol62,Obu62} once the expression \eqref{eq:FAlphaLN} is injected into the Legendre transform \eqref{eq:Legendre}.

We could compute similarly the respective variance of $\ln \varepsilon^\nu_\ell$. It eventually diverges as $|\ell|\to 0$ because finite Reynolds numbers $\mathcal R_e$ \eqref{eq:DefRe} have not been yet taken into account. Before introducing the finite-$\mathcal R_e$ generalization of the probabilistic description of the coarse-grained dissipation \eqref{eq:ProbCGDissInertial}, with the associated distribution of the local exponents \eqref{eq:ProbAlphaCGDissInertial}, we need first to recall that, in this context, the Kolmogorov length scale $\eta_K$ \eqref{eq:EtaK} must be considered as an average behavior of a genuinely fluctuating dissipative length scale  \cite{PalVul87}, according to the parametrization
\begin{align}\label{eq:DefEtaAlpha}
	\eta(\alpha)=L\left( \frac{\mathcal R_e}{\mathcal R^*}\right)^{-\frac{3}{3+\alpha}},
\end{align}
where $L$ is the integral length scale, $\mathcal R^*$ is a universal constant obtained in an empirical way \cite{CheCas05,CheCas06,CheCas12}, related to the so-called Kolmogorov's constant, that is of little importance for the asymptotic behavior we will be seeking, and $\alpha$ a fluctuating exponent, similar to the one entering in \eqref{eq:ProbCGDissInertial}. The dissipative length scale $\eta(\alpha)$ \eqref{eq:DefEtaAlpha} fluctuates around an average value close to the Kolmogorov length scale $\eta_K$ \eqref{eq:EtaK}, which is obtained for the particular value $\alpha=1$, in a range $[\eta_{\min},\eta_{\max}]$ set by the smallest $\alpha_{\min}$ and largest $\alpha_{\max}$ exponents. The probabilistic parametrization for scales $|\ell|<\eta_{\min}$ will then read, once again nicely reviewed in \cite{Bor93},
\begin{align}\label{eq:ProbCGDissDiss}
	\varepsilon^\nu_\ell =\overline{\varepsilon}\left(\frac{\eta(\alpha)}{L}\right)^{\alpha-1} = \overline{\varepsilon}\left( \frac{\mathcal R_e}{\mathcal R^*}\right)^{-\frac{3(\alpha-1)}{3+\alpha}},
\end{align}
with
\begin{align}\label{eq:ProbAlphaCGDissDiss}
	\mathcal F_\ell (\alpha)=\frac{1}{\mathcal Z(0)}\left(\frac{\eta(\alpha)}{L}\right)^{1-f(\alpha)}=\frac{1}{\mathcal Z(0)}\left( \frac{\mathcal R_e}{\mathcal R^*}\right)^{-\frac{3(1-f(\alpha))}{3+\alpha}},
\end{align}
and $\mathcal Z(0)$ the proper renormalization constant such that the integral over $[\alpha_{\min},\alpha_{\max}]$ of \eqref{eq:ProbAlphaCGDissDiss} is equal to unity. With the MF prescriptions given in \eqref{eq:ProbCGDissDiss} and \eqref{eq:ProbAlphaCGDissDiss}, we are now in position to compute the variance $\mathbb V$ of the logarithm of dissipation $\ln \varepsilon^\nu$ according to
\begin{align}\label{eq:ComputVarLogMF}
	\mathbb V\left[ \ln \varepsilon^\nu\right]=\E\left[ \ln^2 \varepsilon^\nu\right]-\E\left[ \ln \varepsilon^\nu\right]^2&=\lim_{|\ell|\to 0}\left[\E\left[ \ln^2 \varepsilon^\nu_\ell\right]-\E\left[ \ln \varepsilon^\nu_\ell\right]^2\right]\notag\\
	&=\ln^2\left( \frac{\mathcal R_e}{\mathcal R^*}\right)\mathbb V \left( \frac{3(\alpha-1)}{3+\alpha}\right),
\end{align}
with
\begin{align}\label{eq:ComputVarLogMF2}
	\mathbb V \left( \frac{3(\alpha-1)}{3+\alpha}\right)=\frac{1}{\mathcal Z(0)}\int_{\alpha_{\min}}^{\alpha_{\max}}\frac{9(\alpha-1)^2}{(3+\alpha)^2}\left( \frac{\mathcal R_e}{\mathcal R^*}\right)^{-\frac{3(1-f(\alpha))}{3+\alpha}}d\alpha-\left(\frac{1}{\mathcal Z(0)}\int_{\alpha_{\min}}^{\alpha_{\max}}\frac{3(\alpha-1)}{3+\alpha}\left( \frac{\mathcal R_e}{\mathcal R^*}\right)^{-\frac{3(1-f(\alpha))}{3+\alpha}}d\alpha\right)^2,
\end{align}
and
\begin{align}\label{eq:Z0MF}
	\mathcal Z(0)=\int_{\alpha_{\min}}^{\alpha_{\max}}\left( \frac{\mathcal R_e}{\mathcal R^*}\right)^{-\frac{3(1-f(\alpha))}{3+\alpha}}d\alpha.
\end{align}

The asymptotic behavior of \eqref{eq:ComputVarLogMF2} heavily relies on the computation of asymptotic behavior of integrals of the type

	$$ I_\phi(\lambda)=\int_{\alpha_{min}}^{\alpha_{max}} \phi(\alpha) e^{\lambda S(\alpha)} d\alpha. $$
	
	where $\lambda$ will be taken as $\ln \frac{\mathcal R_e}{\mathcal R^*}$, $S(\alpha)=- \frac{3}{\alpha+3} (1-f(\alpha) )$ and $\phi$ will be either identically $1$ when computing $\mathcal{Z}$, $g(\alpha)=  3(1-\alpha)/(\alpha+3)$ or $g^2(\alpha)$. With these notations, one can rewrite
		\begin{equation}\label{eq:VarLogMFasymp}
		\mathbb{V}\left( \frac{3(\alpha-1)}{3+\alpha}\right)=    \dfrac{I_{g^2}(\ln \mathcal R_e/\mathcal R^* )}{I_{1}(\ln \mathcal R_e /\mathcal R^*)}- \left(\dfrac{I_{g}(\ln \mathcal R_e/\mathcal R^*)}{I_{1}(\ln\mathcal R_e/\mathcal R^* )} \right)^2. 
	\end{equation}
	As we did previously, the asymptotic behavior of \eqref{eq:VarLogMFasymp} can be derived using Laplace's method. In the present case the first order term of Laplace's method vanishes because \eqref{eq:ComputVarLogMF2} is defined as a difference of two terms. The asymptotic behavior is then obtained by using Laplace's method at the next order in $\lambda$. Since it is a more involved calculation than the lowest order one, we sketch the computation here. The next order in the asymptotic is given by \cite{Won01}
	\begin{equation}\label{eq:LaplaceOrder2}
		I_\phi(\lambda)\underset{\lambda \to \infty}{=} e^{\lambda S(\alpha^\star)} \sqrt{\dfrac{2\pi}{ \lambda |S^{''}(\alpha^\star)|}}\left[ \phi(\alpha^\star)+ \dfrac{C_\phi}{\lambda} + o(\lambda^{-1}) \right],
	\end{equation}
	where $\alpha^\star = \underset{\alpha\in [\alpha_{min}, \alpha_{max}]}{\mathrm{argmax}} S(\alpha)$. Writing $\phi_n= \phi^{(n)}(\alpha^\star)$ and $S_n= S^{(n)}(\alpha^\star)$ the $n-$th derivative of $\phi$ and $S$, the constant $C_\phi$ is given by 
	\begin{equation}\label{eq:ConstOrder2Expansion}
		C_\phi= - \dfrac{ \phi_2}{2S_2} +\phi_1\dfrac{S_3}{2S_2^2}+ \phi_0\dfrac{S_4}{8S_2^2}- \dfrac{5\phi_0}{24} \dfrac{S_3^2}{S_2^3}.
	\end{equation}
	Collecting the lowest non vanishing order terms, one obtains
	$$ \mathbb{V}\left( \frac{3(\alpha-1)}{3+\alpha}\right)  \underset{\mathcal R_e \to \infty}{\sim}  \left[C_{g^2}-2g_0 C_g +g_0^2C_1 \right] \frac{1}{\ln\frac{\mathcal R_e}{\mathcal R^*}} $$
	Using \eqref{eq:ConstOrder2Expansion}, $\alpha^\star= 1+\mu/2$ for the lognormal model and the derivatives of $S$, $g$ and $g^2$ evaluated at $\alpha^\star$ we obtain
	$$ C_{g^2}-2g_0 C_g +g_0^2C_1 = -\dfrac{(g_1)^2}{S_2} =\dfrac{3 \mu}{4} \dfrac{(1+\mu/4)^2}{(1+\mu/8)^3}.$$
	Collecting all the terms together, this yields $ 	\mathbb V\left[ \ln \varepsilon^\nu\right] \sim \dfrac{3 \mu}{4} \dfrac{(1+\mu/4)^2}{(1+\mu/8)^3}\ln\frac{\mathcal R_e}{\mathcal R^*}$. This achieves the justification of the asymptotic \eqref{eq:EstimVarLogMF}.

\section{Logarithmic behavior of the correlation of $X^\epsilon$ in space} \label{App:CorrLog}

The purpose of the present Appendix is the derivation of the logarithmic divergence of the variance of $X^\epsilon$ with $\epsilon$ \eqref {eq:VarXEpsilonFromExKernel}, and the logarithmic behavior of the associated correlation function in the limit $\epsilon\to 0$ \eqref{eq:ResCXEll}, for a particular choice of the convolution kernel given in \eqref{eq:ExampleFTGLEpsilonText}, that we repeat here for convenience
 \begin{align}\label{eq:ExampleFTGLEpsilon} 
\widehat{G}_L^{\epsilon}(k) =\frac{1}{\sqrt{A_{d-1}}}\frac{1}{|k|^{d/2}}\mathds{1}_{1/L\le |k|\le 1/\epsilon},
\end{align}
where is understood that $\epsilon<L$. 

As a matter of fact, the variance of $X^\epsilon$ reads
\begin{align}\label{eq:VarXEpsilon} 
\E\left[\left(X^\epsilon\right)^2\right] &=\int_{\R^d}\left|\widehat{G}_L^{\epsilon}(k) \right|^2dk=\int_{L^{-1}}^{\epsilon^{-1}}\frac{1}{\rho}d\rho=\ln\left( \frac{L}{\epsilon}\right).
\end{align}

In the limit $\epsilon\to 0$, the covariance of $X^\epsilon$ \eqref{eq:CovXEpsilonFromWiener}  becomes
 \begin{align}\label{eq:LimExampleCovXEpsilon1} 
\mathcal C_{X}(|\ell|)=\lim_{\epsilon \to 0}\E\left[X^\epsilon(x)X^\epsilon(x+\ell)\right] &= \frac{1}{A_{d-1}}\int_{|k|\ge 1/L}e^{2i\pi k\cdot \ell} \frac{dk}{|k|^d}.
\end{align}
 This is a classical integral that enters in the analysis of the Fourier transform of homogeneous functions, in particular of critical order \cite{Hor15,SteWei71}, that can be rigorously shown to behave logarithmically at small arguments. To see this, we will be using standard technical steps nicely reviewed in \cite{Gra08} (see in particular the appendix B devoted to Bessel functions and Fourier transform of radial functions). Following these steps, introducing the Bessel function of the first kind $J_\nu(x)$, we get
 \begin{align}\label{eq:LimExampleCovXEpsilon2} 
\mathcal C_{X}(|\ell|)&=\frac{2\pi|\ell|^{1-d/2}}{A_{d-1}}\int_{\rho\ge 1/L}\rho^{-d/2}J_{d/2-1}(2\pi|\ell|\rho)d\rho\notag\\
&=\frac{(2\pi)^{d/2}}{A_{d-1}}\int_{s\ge 2\pi|\ell|/L}s^{-d/2}J_{d/2-1}(s)ds,
\end{align}
with the consequence that $\mathcal C_{X}$ is indeed an isotropic function, and solely depends on the norm $|\ell|$. Then split the integral over the intervals $2\pi|\ell|/L \le s\le 1$ and $s\ge 1$. To deal with the first integral, we make use of the behavior of the Bessel functions at small arguments, as it is conveniently recalled in \cite{Gra08}, that reads
\begin{align}\label{eq:DevBessSmallArg} 
J_\nu(s) = \frac{s^\nu}{2^\nu\Gamma(\nu+1)} + S_\nu(s),
\end{align}
where $|S_\nu(s)|\le \text{cst}\,s^{\nu+1}$, such that
\begin{align}\label{eq:FirstTermLimExampleCovXEpsilon} 
\int_{2\pi|\ell|/L}^1s^{-d/2}J_{d/2-1}(s)ds = \frac{1}{2^{d/2-1}\Gamma(d/2)}\ln\left( \frac{L}{2\pi |\ell|}\right)+\int_{2\pi|\ell|/L}^1s^{-d/2}S_{d/2-1}(s)ds,
\end{align}
the second term entering in the RHS of \eqref{eq:FirstTermLimExampleCovXEpsilon}  being bounded with $|\ell|$. Again, as proposed in \cite{Gra08}, make use of the following identity, useful at large arguments,
\begin{align}\label{eq:DevBessLargeArg} 
J_\nu(s) = \sqrt{\frac{2}{\pi s}}\cos\left( s-\frac{\pi \nu}{2}-\frac{\pi}{4}\right) + R_\nu(s),
\end{align}
with $|R_\nu(s)|\le C_\nu\,s^{-3/2}$ for any $s\ge1$, to conclude that the remaining integral entering in \eqref{eq:LimExampleCovXEpsilon2} over the range  $s\ge 1$ is bounded with $|\ell |$ (for the special case $d=1$, make an additional integration by parts). As a conclusion, we easily see, using \eqref{eq:FirstTermLimExampleCovXEpsilon} that the covariance of $X^\epsilon$ in the limit $\epsilon\to 0$ \eqref{eq:LimExampleCovXEpsilon1} diverges logarithmically at the origin, the overall behavior at any $|\ell|$ and finite $\epsilon>0$ being nicely described by the functional form depicted in \eqref{eq:DefCorrXEpsilon}. 

Accordingly, the remaining continuous and bounded function $f_X$ entering in the asymptotic expression of the covariance function \eqref{eq:ResCXEll} reads
\begin{equation}\label{eq:FXELLSpatial}
	f_X(\ell)=-\ln(2\pi)+
	\begin{cases}
	&\displaystyle\dfrac{(2\pi)^{\frac d2}}{A_{d-1}}\left[    \int_{2\pi |\ell|/L}^1 s^{-d/2}S_{d/2-1}(s)ds+ \int_1^\infty s^{-d/2}J_{d/2-1}(s)ds \right]  \text{ if }2\pi |\ell|\leq L\\
	&\displaystyle \dfrac{(2\pi)^{\frac d2}}{A_{d-1}}  \int_{2\pi |\ell|/L}^\infty s^{-d/2}J_{d/2-1}(s)ds \text{ if }2\pi |\ell|\geq L.
\end{cases}
\end{equation}
Thanks to the behavior of $S_\nu$ at the origin stated previously, the integrand of the first integral is finite at the origin making the function $f$ a  continuous function of its argument. In particular, using a symbolic calculation software, we find that 
\begin{equation}\label{eq:ValueFXAtOrigin}
	\lim_{|\ell|\to 0} f_X(\ell) =f_{X}(0)= -\ln \pi + \dfrac{\psi(d/2)-\varrho}{2}=
	\begin{cases}
       &-\ln(2\pi)-\varrho, ~ \text{if}~d =1\\
       &-\ln \pi-\varrho, ~ \text{if}~d =2\\
       &1-\ln(2\pi)-\varrho, ~\text{if}~d =3,
	\end{cases}
\end{equation}
where $\psi$ is the digamma function and $\varrho \simeq 0.577$ the Euler-Mascheroni constant.

\section{Logarithmic behavior of the spatio-temporal correlation of $X^\epsilon(t,x)$ } \label{App:CorrLogST}

Similarly to the former Appendix \ref{App:CorrLog}, we would like now to show the proposed expression \eqref{eq:ResCXTauEllMax} for the covariance function $\mathcal C_{X_\beta}(\tau,\ell)$ \eqref{eq:CXBetaAsympt} of the spatio-temporal random field $X_\beta^\epsilon(t,x)$ in the asymptotic limit $\epsilon\to 0$ for the particular choice of the deterministic kernel given in \eqref{eq:ExampleFTGLEpsilon}.

Accordingly, we begin with
\begin{align}\label{eq:LimExampleCovXEpsilon1ST} 
\mathcal C_{X_\beta}(\tau,\ell) &= \frac{1}{A_{d-1}}\int_{|k|\ge 1/L}e^{2i\pi k\cdot \ell} e^{-D_3|\tau| |k|^{2\beta}}\frac{dk}{|k|^d}.
\end{align}

Letting $R= \max\left(2\pi |\ell|, (D_3 |\tau|)^{\frac{1}{2\beta}} \right)$, we get 
 \begin{align*} 
	C_{X_\beta}(\tau,\ell)&=\frac{2\pi|\ell|^{1-d/2}}{A_{d-1}}\int_{\rho\ge 1/L}\rho^{-d/2}e^{-D_3|\tau| \rho^{2\beta}}J_{d/2-1}(2\pi|\ell|\rho)d\rho\\
	&=\frac{2\pi}{A_{d-1}} \left(\dfrac{|\ell|}{R}\right) ^{1-d/2}\int_{R/L}^\infty s^{-d/2}e^{-\frac{D_3|\tau|}{R^{2\beta}}  s^{2\beta}}J_{d/2-1}\left(\dfrac{2\pi |\ell|}{R}s \right)ds.\\
\end{align*}
Then split the integral over the intervals $R/L \le s\le 1$ and $1 \le s$. Once again, the first interval will lead to the logarithmic behavior. Using \eqref{eq:DevBessSmallArg}, we obtain
 \begin{multline} \label{eq:SplitIntegral1}
\frac{2\pi}{A_{d-1}} \left(\dfrac{|\ell|}{R}\right) ^{1-d/2}	\int_{R/L}^1 s^{-d/2}e^{-\frac{D_3|\tau| s^{2\beta}}{R^{2\beta}}}J_{d/2-1}\left(\frac{2\pi |\ell|}{R}s \right)ds= \ln \left( \frac{L}{R} \right)+ \int_{R/L}^1 \frac{e^{-\frac{D_3|\tau|}{R^{2\beta}} s^{2\beta} }-1}{s}ds\\+\frac{2\pi}{A_{d-1}} \left(\dfrac{|\ell|}{R}\right) ^{1-d/2}\int_{R/L}^1 s^{-d/2}e^{-\frac{D_3|\tau| s^{2\beta}}{R^{2\beta}}}S_{d/2-1}\left(\frac{2\pi |\ell|}{R}s \right)ds.
\end{multline}
We can therefore write
\begin{equation}
	\mathcal C_{X_\beta}(\tau,\ell)= \ln_+\left( \dfrac{L}{ \max\left(2\pi |\ell|, (D_3 |\tau|)^{\frac{1}{2\beta}} \right)} \right)+f_{X_\beta}(\tau,\ell).
\end{equation}
When $R\geq L$, we have $f_{X_\beta}(\tau,\ell)=\mathcal C_{X_\beta}(\tau,\ell)$, that goes to zero for large $R$. For $R\leq L$, we have using \eqref{eq:SplitIntegral1}
\begin{multline} \label{eq:SplitIntegral2}
	f_{X_\beta}(\tau,\ell)= \int_{R/L}^1 \frac{e^{-\frac{D_3|\tau|}{R^{2\beta}} s^{2\beta} }-1}{s}ds+\frac{2\pi }{A_{d-1}}  \left(\dfrac{|\ell|}{R}\right) ^{1-d/2}\int_{R/L}^1 s^{-d/2}e^{-\frac{D_3|\tau| s^{2\beta}}{R^{2\beta}}}S_{d/2-1}\left(\frac{2\pi |\ell|}{R}s \right)ds\\+\frac{2\pi}{A_{d-1}}  \left(\dfrac{|\ell|}{R}\right) ^{1-d/2}\int_{1}^\infty s^{-d/2}e^{-\frac{D_3|\tau| s^{2\beta}}{R^{2\beta}}}J_{d/2-1}\left(\dfrac{2\pi |\ell|}{R}s \right)ds .
\end{multline}
Let us now show that $f_{X_\beta}(\tau,\ell)$ \eqref{eq:SplitIntegral2} is bounded for $R\leq L$. The first integral in \eqref{eq:SplitIntegral2} is bounded because for $x\geq 0$ we have $|e^{-x}-1|\leq x$, such that 
$$\left|\int_{R/L}^1 \frac{e^{-\frac{D_3|\tau|}{R^{2\beta}} s^{2\beta} }-1}{s}ds \right| \leq \dfrac{D_3|\tau|}{R^{2\beta}} \dfrac{1-(R/L)^{2\beta}}{2\beta}\leq \dfrac{1}{2\beta}. $$
Concerning the second integral of \eqref{eq:SplitIntegral2}, we bound the exponential by 1, and make use of $ |S_\nu(x)|\leq c_\nu x^{\nu+1}$ with $c_\nu>0$ \cite{Gra08}, in order to obtain
\begin{equation*}
\left|\frac{2\pi}{A_{d-1}} \left(\dfrac{|\ell|}{R}\right) ^{1-d/2}\int_{R/L}^1 s^{-d/2}e^{-\frac{D_3|\tau| s^{2\beta}}{R^{2\beta}}}S_{d/2-1}\left(\frac{2\pi |\ell|}{R}s \right)ds \right| \leq \dfrac{c_{d/2-1}(2\pi)^{d/2}}{A_{d-1}} \dfrac{2\pi |\ell|}{R}\left( 1- \dfrac{R}{L}\right)\leq \dfrac{c_{d/2-1}(2\pi)^{d/2}}{A_{d-1}}.
\end{equation*}
For the convergence of the last integral in \eqref{eq:SplitIntegral2}, we have to consider two cases. Firstly, when $R=2\pi |\ell|$, we bound the exponential by one and use a similar argument as developed to show boundedness of  \eqref{eq:FXELLSpatial}. When $R= (D_3|\tau|)^{1/(2\beta)}$, we can rewrite the integral as 
\begin{equation}
	\frac{2\pi}{A_{d-1}} \left(\dfrac{|\ell|}{R}\right) ^{1-d/2}\int_{1}^\infty s^{-d/2}e^{- s^{2\beta}}J_{d/2-1}\left(\frac{2\pi |\ell|}{R}s \right)ds=\frac{2\pi}{A_{d-1}} \int_{1}^\infty s^{-1}e^{- s^{2\beta}}H_{d/2-1}\left(\frac{2\pi |\ell|}{R}s \right)ds
\end{equation}
where $H_\nu(s)= s^{-\nu} J_\nu(s)$ which is bounded on the whole real line. Because of the boundedness of $H_{d/2-1}$, the integral is finite thanks to the exponential cutoff.

Due to the different conventions used for purely spatial correlations \eqref{eq:ResCXEll} and spatio-temporal ones \eqref{eq:ResCXTauEllMax}, the limiting value of $f_{X_\beta}(0,\ell)$ is given by \eqref{eq:ValueFXAtOrigin} with and additional factor $\ln 2\pi$. Let us compute the value at the origin of its temporal counterpart $f_{X_\beta}(\tau,0)$ in order to comment on the continuity of the function of two variables $f_{X_\beta}(\tau,\ell)$ in the vicinity of the origin \eqref{eq:SplitIntegral2}. We have 
\begin{align}\label{eq:TimeLineST}
\mathcal C_{X_\beta}(\tau,0)= \ln\dfrac{L}{(D_3 |\tau|)^{\frac{1}{2\beta}}}+ \int_{(D_3 |\tau|)^{\frac{1}{2\beta}}}^1 \dfrac{e^{-s^{2\beta}}-1}{s}ds+ \int_1^\infty \dfrac{e^{-s^{2\beta}}}{s}ds, 
\end{align}
and using a symbolic calculation software, we obtain
\begin{align} \label{eq:DefFXTemp}
 f_{X_\beta}^{\text{\tiny{temp}}} \equiv \lim_{\tau\to 0}f_{X_\beta}(\tau,0)= \int_{0}^1 \dfrac{e^{-s^{2\beta}}-1}{s}ds+ \int_1^\infty \dfrac{e^{-s^{2\beta}}}{s}ds=-\dfrac{\varrho}{2\beta}. 
 \end{align} 
Recall that we have computed in the purely spatial case the value 
\begin{align} \label{eq:DefFXSpace}
 f_{X_\beta}^{\text{\tiny{space}}} \equiv \lim_{|\ell|\to 0}f_{X_\beta}(0,\ell),
 \end{align} 
that was found to depend on the spatial dimension $d$ \eqref{eq:ValueFXAtOrigin}. Thus, when compared to \eqref{eq:DefFXTemp}, we are led to the conclusion that
\begin{align} \label{eq:FXSpaceFXTempNotEqual}
 f_{X_\beta}^{\text{\tiny{space}}} \ne  f_{X_\beta}^{\text{\tiny{temp}}} .
 \end{align} 
This implies the discontinuity of the function of two variables  $f_{X_\beta}(\tau,\ell)$ in the vicinity of the origin \eqref{eq:SplitIntegral2}.

\section{On the periodic version of the model} \label{App:PeriodicModel}

Let us explain some details concerning the adaptation of the model \eqref{eq:OUFTX} on the one dimensional torus of length $\Lt$. For $k\in \mathbb{Z}/\Lt$ the solution of the model writes,
$$ \widehat{X}_\beta^\epsilon(t,k) = \widehat{G}^\epsilon_L(k)\sqrt{\dfrac{2}{T_{k,\beta}}}   \int_{-\infty}^t e^{- (t-s)/T_{k,\beta}} d\widehat{W}(s,k).$$
 This is the same overall expression than in the case $k\in \mathbb{R}$ but here $k$ takes discrete values and the white noise is of vanishing average and covariance
 \begin{equation}\label{eq:CovWhiteNoiz}
 	\mathbb{E}\left[ d\widehat{W}(t_1,k_1)\overline{d\widehat{W}(t_2,k_2)}\right]
 	=\Lt\,\delta_{k_1,k_2}\delta(t_1-t_2)\,dt_1\,dt_2,
 \end{equation}
 where $\overline{\cdot}$ denotes complex conjugation, $\delta_{k_1,k_2}$ the Kronecker delta, and $\delta(t)$ the Dirac distribution. See \cite{ChaLem26} for additional discussions concerning the expression of these type of models on the torus. The inverse Fourier series writes
 $$X_\beta^\epsilon(t,x)= \dfrac{1}{\Lt}\sum_{ n \in  \mathbb{Z}}e^{2i\pi k_n x}  \widehat{G}^\epsilon_L(k_n) \sqrt{\dfrac{2}{T_{k_n,\beta}}}  \int_{-\infty}^t e^{- (t-s)/T_{k_n,\beta}} d\widehat{W}(s,k_n), $$
 where $k_n=n/\Lt$. Using \eqref{eq:CovWhiteNoiz}, together with $\widehat{G}^\epsilon_L(k_n)= 1/\sqrt{2|k_n|} \mathds{1}_{1/L\le |k_n|\le 1/\epsilon}$ and $T_{k_n,\beta=1/2}$ it is straightforward to check that 
 $$	\mathbb{E}\left[ X_\beta^\epsilon(t,x) X_\beta^\epsilon(t+\tau,x+\ell) \right]:= \mathcal C_{X_\beta^\epsilon}(\tau,\ell)= \sum_{n=n_L}^{n_\epsilon}\dfrac{\exp\left(-\frac{D_3 n}{\Lt} |\tau|\right)}{n}\cos\left(\dfrac{2\pi n}{\Lt} \ell\right),  $$
 where $n_L= \lceil \Lt/L\rceil$ and $n_\epsilon =  \lfloor \Lt/\epsilon\rfloor$ and $\lceil \cdot \rceil$ and $\lfloor \cdot \rfloor$ are respectively the ceil and floor functions.
 Let us then, as in the $k-$continuous case, establish the logarithmic correlation structure together with the additional constants entering the correlation at small lags.

\subsection{Spatial correlations}

Let us study the vanishing $\epsilon$ behavior of the spatial correlations,
\begin{equation}\label{eq:CorrSpaceTorus}
	\lim_{\epsilon \to 0}\mathcal C_{X_\beta^\epsilon}(0,\ell)= \sum_{n=n_L}^{\infty}\dfrac{\cos\left(\frac{2\pi n}{\Lt} \ell\right)}{n} .
\end{equation}

In order to establish the logarithmic correlations in space, one need to use the power series expansion of the principal logarithm in the complex plane, that is $\ln(1-z) = - \sum_{n=1}^\infty z^n /n$ for any $|z|<1$. In our case, $z=e^{2i\pi |\ell|/\Lt}$ is of modulus unity breaking down, a priori, the power series expansion. However, note that whenever $\ell/\Lt \notin \mathbb{N}$, the series \eqref{eq:CorrSpaceTorus} converges. Abel's theorem therefore ensure that  the power series expansion can be extended up to the boundary point. This yields
\begin{equation}
	\lim_{\epsilon \to 0}\mathcal C_{X_\beta^\epsilon}(0,\ell)= - \ln \left( 2 \left|\sin\left(\dfrac{\pi \ell}{\Lt}   \right)\right|\right)- \sum_{n=1}^{n_L-1}\dfrac{\cos\left(\frac{2\pi n}{\Lt} \ell\right)}{n}  \underset{\ell \to 0}{\sim}  \ln \left(  \dfrac{ \Lt }{2\pi \ell}   \right)- \sum_{n=1}^{n_L-1}\dfrac{1}{n}.
\end{equation}
Similarly to the study of DNS performed in Section~\ref{Sec:NumObservationsHopkins}, we can absorb the constant in $\ell$ term by defining 
\begin{equation}\label{eq:LstarPeriodic}
	L_{X_\beta}^\star= \dfrac{\Lt}{2\pi} e^{-H_{n_L-1}},
\end{equation}
   where $H_N = \sum_{n=1}^N 1/n= \ln N + \varrho + o_N(1)$ is the Harmonic sum. In the large $\Lt$ asymptotic, we expect to recover $	L_{X_\beta}^\star \to L \exp(f_{X_\beta}^{\text{\tiny space}})=\frac{L}{2\pi} e^{-\varrho}$ obtained from the model on $\R$. Using the asymptotic behavior of the Harmonic sum we indeed obtain,
    
 $$  	L_{X_\beta}^\star \underset{\Lt \to \infty}{\sim} \dfrac{\Lt}{2\pi} e^{ -\ln \frac{\Lt}{L}- \varrho}= \dfrac{L}{2\pi}e^{-\varrho}=L e^{f_{X_\beta}^{\text{\tiny space}}}.$$
The periodic setting thus provide a natural approximation of the field on the real line at the level of spatial correlations. 
   
 \subsection{Temporal correlations}
 Let us now turn to the fully temporal case. We use again the power series expansion of the logarithm but this time for a real variable and away from the boundary. The temporal correlations thus write
 $$ \lim_{\epsilon \to 0 }\mathcal C_{X_\beta^\epsilon}(\tau,0)= -\ln \left( 1-e^{- \frac{D_3}{\Lt}|\tau|}\right)- \sum_{n=1}^{n_L-1} \dfrac{e^{- \frac{D_3 n}{\Lt}|\tau|}}{n}\underset{\tau \to 0}{\sim}   \ln\left( \dfrac{\Lt}{D_3 |\tau|}\right)- \sum_{n=1}^{n_L-1} \dfrac{1}{n}.$$
 This asymptotic behavior allows one to introduce as we did previously with the DNS, the time scale entering the normalization of time lag in the simulations
 
\begin{equation}\label{eq:TstarPeriodic}
 	T_{X_\beta}^\star=\dfrac{\Lt}{D_3}e^{-H_{n_L-1}} 
\end{equation}

Using the asymptotic behavior of the Harmonic sum  we have in addition 
 $$ T_{X_\beta}^\star\underset{\Lt \to \infty}{\sim}  \dfrac{\Lt}{D_3} e^{- \ln\left(\Lt/L\right) - \varrho } = \dfrac{L}{D_3}e^{-\varrho}=\dfrac{L}{D_3}e^{f_{X_\beta}^{\text{\tiny temp}}} .$$
 Therefore, in the infinite domaine size limit, the characteristic time scale $T_{X_\beta}^\star$ matches the one of the model on $\R$.

 \subsection{Numerical scheme}\label{App:NumSchem}

 For the sake of completeness, let us present the numerical scheme adapted from \cite{ChaLem26} that we use to perform exact in law numerical simulations of the dynamics \eqref{eq:OUFTX}. Between two times $t^{p}$ and $t^{p+1}$ distant from $\delta t$ and for some $k\in \mathbb{Z} / \Lt$, we have 
 $$ \widehat{X}_\beta^\epsilon(t^{p+1},k)=e^{- \frac{\delta t}{T_{k,\beta}}}\widehat{X}_\beta^\epsilon(t^{p},k)+\widehat{G}_L^\epsilon(k)\sqrt{ \dfrac{2}{T_{k,\beta}}} \int_{t^p}^{t^{p+1}} e^{- \frac{t^{p+1}-s}{T_{k,\beta}}}d \widehat{W}(s,k).$$
The stochastic integral entering the above expression is a gaussian random variable of vanishing average and variance
$$ \mathbb{E} \left|\int_{t^p}^{t^{p+1}} e^{- \frac{t^{p+1}-s}{T_{k,\beta}}}d \widehat{W}(s,k) \right|^2= \Lt \dfrac{T_{k,\beta}}{2}\left[ 1-e^{-\frac{2\delta t} {T_{k,\beta}} }\right]. $$
This information, together with the delta-correlation in $k$  of the Fourier modes of the white noise, leads to the following exact in law scheme,
$$ \widehat{X}_\beta^\epsilon(t^{p+1},k)=e^{- \frac{\delta t}{T_{k,\beta}}}\widehat{X}_\beta^\epsilon(t^{p},k)+\widehat{G}_L^\epsilon(k)\sqrt{ 1-e^{-\frac{2\delta t} {T_{k,\beta}} }} \gamma^{p}(k).  $$
The random forcing is taken into account by $\gamma^{p}(k)$, complex Gaussian variables of vanishing average and variance $\Lt$. The $\gamma^{p}(k)$'s are i.i.d random variables, up to the Hermitian symmetry $\overline{\gamma^{p}(k)} =\gamma^{p}(-k)$. See \cite{ChaLem26} for a more detailed discussion on this numerical scheme and the extension of the model to smooth in time field.

 \small
\bibliographystyle{unsrt}

\end{document}